\documentclass[journal,twocolumn]{IEEEtran}
\usepackage{amsmath,graphicx}
\usepackage{color}
\usepackage{graphicx}
\usepackage{epstopdf}
\usepackage{amsmath}
\usepackage{amssymb}
\usepackage{tikz}

\usepackage[english]{babel}
\usepackage{cite}
\usepackage{rotfloat}
\usepackage{mathtools}
\usepackage[font=normalsize,labelfont=bf]{caption}
\usepackage{amsmath}
\usepackage{makecell}
\usepackage{algorithm,algorithmic}
\usepackage{multirow}
\usepackage{subfigure}
\usepackage{booktabs}
\usepackage{colortbl}
\usepackage{multirow}
\usepackage{hhline}
\usepackage{stfloats}
\usepackage{multicol}
\usepackage{bbm}
\usepackage{cases}
\newsavebox{\foobox}

\definecolor{kugray5}{RGB}{224,224,224}

\usepackage[normalem]{ulem}
\newcommand\rsout{\bgroup\markoverwith
	{\textcolor{red}{\rule[0.5ex]{2pt}{0.8pt}}}\ULon}



\makeatletter
\newcommand{\ALOOP}[1]{\ALC@it\algorithmicloop\ #1%
	\begin{ALC@loop}}
	\newcommand{\ENDALOOP}{\end{ALC@loop}\ALC@it\algorithmicendloop}

\makeatother

\usepackage{etoolbox}
\let\mybibitem\bibitem
\renewcommand{\bibitem}[1]{%
	\ifstrequal{#1}{nature}
	{\color{blue}\mybibitem{#1}}
	{\color{black}\mybibitem{#1}}%
}

\graphicspath{ {Figures/} }

\DeclareCaptionLabelSeparator{periodspace}{.\quad}

\addto\captionsenglish{}
\newcommand\nbthis{\addtocounter{equation}{1}\tag{\theequation}}

\newcommand{\norm}[1]{\left\lVert#1\right\rVert} 
\newcommand{\normshort}[1]{\lVert#1\rVert} 
\newcommand{\abs}[1]{\left|#1\right|} 
\newcommand{\absshort}[1]{|#1|} 

\newcommand{\tr}[1]{\mathrm{trace}\left(#1\right)} 

\newcommand{\re}[1]{\mathfrak{R}{\left(#1\right)}}
\newcommand{\im}[1]{\mathfrak{I}{\left(#1\right)}}
\allowdisplaybreaks

\newcommand{\mean}[1]{\mathbb{E} \left\{#1\right\}}

\usepackage{setspace}

\newcommand{\mQ}{{\mathbf{Q}}}

\newcommand{\mH}{{\mathbf{H}}} 

\newcommand{\mA}{{\mathbf{A}}}

\newcommand{\mI}{\textbf{\textbf{I}}}

\newcommand{\mB}{{\mathbf{B}}}
\newcommand{\mC}{{\mathbf{C}}}

\newcommand{\mF}{{\mathbf{F}}}
\newcommand{\mU}{{\mathbf{U}}}
\newcommand{\mV}{{\mathbf{V}}}

\newcommand{\setC}{\mathbb{C}} 
\newcommand{\setR}{\mathbb{R}}

\newcommand{\setN}{\mathbb{N}}

\newcommand{\vc}{{\mathbf{c}}}

\newcommand{\vx}{{\mathbf{x}}}
\newcommand{\vy}{{\mathbf{y}}}

\newcommand{\vu}{{\mathbf{u}}}
\newcommand{\vz}{{\mathbf{z}}}

\newcommand{\vw}{{\mathbf{w}}}
\newcommand{\va}{{\mathbf{a}}}

\newcommand{\vf}{\mathbf{f}}

\def\argmin{\mathop{\mathrm{argmin}}}

\def\b0{{\pmb{0}}} 

\def\ba{{\mathbf{a}}}

 \def\bn{{\mathbf{n}}}  
  \def\bs{{\mathbf{s}}} 
   \def\bx{{\mathbf{x}}}

 \def\bB{{\mathbf{B}}}  
 \def\bF{{\mathbf{F}}}  \def\bH{{\mathbf{H}}}
\def\bI{{\mathbf{I}}}

 \def\bV{{\mathbf{V}}}

\newcommand{\Frf}{\bF_\text{RF}}
\newcommand{\Fopt}{\mF_{\text{opt}}}

\newcommand{\Frft}{\tilde{\bF}_\text{RF}}
\newcommand{\Fbb}{\bF_\text{BB}}

\newcommand{\Nrf}{N_{\text{RF}}} 
\newcommand{\Nr}{N_\text{r}}
\newcommand{\Nt}{N_\text{t}}

\newcommand{\Ns}{N_\text{s}}
\newcommand{\Nrh}{N_\text{r}^\text{h}}
\newcommand{\Nrv}{N_\text{r}^\text{v}}
\newcommand{\Nth}{N_\text{t}^\text{h}}
\newcommand{\Ntv}{N_\text{t}^\text{v}}

\newcommand{\ar}{\ba_\text{r}} 
\newcommand{\at}{\ba_\text{t}}

\newcommand{\Afull}{\mathcal{A}_{\text{full}}}
\newcommand{\Asub}{\mathcal{A}_{\text{sub}}}

\begin{document}
	\setlength{\abovedisplayskip}{4.5pt}
	\setlength{\belowdisplayskip}{4.5pt}
	\title{Deep Unfolding Hybrid Beamforming Designs for THz Massive MIMO Systems}
	\author{Nhan Thanh Nguyen, \IEEEmembership{Member, IEEE},
		Mengyuan Ma, \IEEEmembership{Student Member, IEEE}, Ortal Lavi,
		Nir Shlezinger, \IEEEmembership{Member, IEEE},
		Yonina C. Eldar, \IEEEmembership{Fellow, IEEE},
		A.~L.~Swindlehurst, \IEEEmembership{Fellow, IEEE},
		and Markku Juntti, \IEEEmembership{Fellow, IEEE}
		\thanks{This research was supported by Academy of Finland under 6Genesis Flagship (grant 318927), EERA Project (grant 332362), Infotech Program funded by University of Oulu Graduate School, and U.S. National Science Foundation grant CCF-2225575. The authors wish to acknowledge CSC – IT Center for Science, Finland, for computational resources. A short version of this paper  has been submitted to the IEEE Int. Conf. Acoust., Speech, Signal Processing, 2023.}
		\thanks{N. T. Nguyen, M. Ma, and M. Juntti are with Centre for Wireless Communications, University of Oulu, P.O.Box 4500, FI-90014, Finland (e-mail:
			\{nhan.nguyen, mengyuan.ma, markku.juntti\}@oulu.fi). N. Shlezinger and Ortal Lavi are with School of ECE, Ben-Gurion University of the Negev, Beer-Sheva, Israel (email: \{nirshl, agivo\}@bgu.ac.il). Y. C. Eldar is with Faculty of Math and CS, Weizmann Institute of Science, Rehovot, Israel (email: yonina.eldar@weizmann.ac.il). A. L. Swindlehurst is with Department of EECS, University of California, Irvine, CA, US (email: swindle@uci.edu).}
	}
	
	\maketitle
	
	\begin{abstract}
		Hybrid beamforming (HBF) is a key enabler for wideband terahertz (THz) massive multiple-input multiple-output (mMIMO) communications systems. A core challenge with designing HBF systems stems from the fact their application often involves a non-convex, highly complex optimization of large dimensions. In this paper, we propose  HBF schemes that leverage data to enable efficient designs for both the fully-connected HBF (FC-HBF) and dynamic sub-connected HBF (SC-HBF) architectures. We develop a deep unfolding framework based on factorizing the optimal fully digital beamformer into analog and digital terms and formulating two corresponding equivalent least squares (LS) problems. Then, the digital beamformer is obtained via a closed-form LS solution, while the analog beamformer is obtained via ManNet, a lightweight sparsely-connected deep neural network based on unfolding projected gradient descent. Incorporating ManNet into the developed deep unfolding framework leads to the ManNet-based FC-HBF scheme. We show that the proposed ManNet can also be applied to SC-HBF designs after determining the connections between the radio frequency chain and antennas. We further develop a simplified version of ManNet, referred to as subManNet, that directly produces the sparse analog precoder for SC-HBF architectures. Both networks are trained with an unsupervised training procedure. Numerical results verify that the proposed ManNet/subManNet-based HBF approaches outperform the conventional model-based and deep unfolded counterparts with very low complexity and a fast run time. For example, in a simulation with $128$ transmit antennas, it attains a slightly higher spectral efficiency than the Riemannian manifold scheme, but over $1000$ times faster and with a complexity reduction of more than by a factor of six (6).
		
	\end{abstract}
	
	\begin{IEEEkeywords}
		THz communications, hybrid beamforming, massive MIMO, deep learning, AI, deep unfolding.
	\end{IEEEkeywords}
	\IEEEpeerreviewmaketitle
	
	\section{Introduction}
	\label{sec:intro}
	
	Future sixth-generation (6G) wireless networks are expected to realize Tbps single-user data rates to support emerging ultra-high-speed applications, such as mobile holograms, immersive virtual reality, and digital twins \cite{giordani2020toward}. To realize such rapid growth in data traffic and applications, wideband terahertz (THz) massive multiple-input multiple-output (mMIMO) systems have emerged as key enablers for achieving substantial improvements in the system spectral and energy efficiency (SE/EE) \cite{rappaport2019wireless}. In THz mMIMO transceivers, hybrid beamforming (HBF) can provide a cost- and energy-efficient solution that yields significant multiplexing gains with a limited number of power-hungry radio frequency (RF) chains \cite{gao2018low, dai2022delay}. 
	
	As HBF delegates some of the beamforming operations to the analog domain, its design largely depends on the considered hardware and its associated constraints~\cite{ioushua2019family}. A candidate implementation of HBF systems realizes analog beamforming via tunable complex gains and phase shifters~\cite{gong2020rf}, which can be efficiently designed using quantized vector modulators~\cite{tasci2022robust}. While these architectures are highly flexible, they are expected to be very costly when implemented at high frequencies. Another candidate HBF architecture is based on metasurface antennas~\cite{shlezinger2021dynamic}, whose implementation for mMIMO at high frequencies is still an area of active research. Consequently, the most common mMIMO HBF architecture considered to date realizes analog beamforming using adjustable phase shifters~\cite{mendez2016hybrid}. 
	However, optimizing a phase-shifter-based HBF is challenging due to the need for constant modulus constraints on the analog beamforming coefficients and the strong coupling between the analog and digital beamformers. Thus, efficient HBF methods overcoming these challenges have attracted much interest in the literature, with proposed approaches ranging from conventional model-based optimizations to purely data-driven deep learning (DL). 
	
	\subsection{Related Works}
	HBF designs and optimization usually require cumbersome algorithms such as Riemannian manifold minimization (MO-AltMin) \cite{yu2016alternating} and alternating optimization (AO) \cite{sohrabi2016hybrid}. In MO-AltMin, the alternating analog and digital beamformer designs form a nested loop procedure, wherein the former is solved by Riemannian manifold optimization, and the latter is obtained via a least squares (LS) problem. With $\Nt$ antennas and $\Nrf$ RF chains, AO solves for each of $\Nt \Nrf$ analog beamforming coefficients in an alternating manner until convergence. Although MO-AltMin and AO offer satisfactory performance, both require nested loops with high complexity and slow convergence, especially for large mMIMO systems. A low-complexity alternative for HBF designs is the orthogonal matching pursuit (OMP) approach \cite{el2014spatially}. It requires only $\Nrf$ iterations to select $\Nrf$ analog precoding vectors from a codebook consisting of array response vectors. However, the performance of OMP is usually significantly inferior to the optimum. 
	
	While MO-AltMin works for both narrowband and wideband scenarios, the original AO and OMP approaches only apply to narrowband systems. Lee \textit{et al.} \cite{lee2014af} further optimized OMP for orthogonal frequency-division multiplexing (OFDM)-based MIMO systems. In \cite{nguyen2022beam}, a variant of AO was proposed for wideband MIMO-OFDM systems. It is shown that an analog combiner designed only for the center frequency and optimal frequency-dependent digital combiners can achieve near-optimal performance as long as the bandwidth is narrow or the array's dimensions are small enough so that the array response remains approximately frequency-non-selective. When the array response becomes \emph{frequency-selective} or suffers from the so-called \emph{beam squint} effect \cite{nguyen2022beam} in wideband THz systems, it can be mitigated by employing true-time-delay (TTD) lines in the analog beamforming architecture \cite{dai2022delay, tan2019delay, dovelos2021channel}. However, the deployment of TTDs requires additional hardware complexity and power consumption. Yuan \textit{et al.}  \cite{yuan2020hybrid} proposed a  wideband HBF scheme with two digital beamformers, in which an additional digital beamformer is introduced to compensate for the performance loss caused by the constant-amplitude hardware constraints and channel non-uniformity across the subcarriers. Li \textit{et al.} \cite{li2020dynamic} considered an HBF architecture with dynamic antenna subarrays and low-resolution phase shifters and address the HBF design with classical block coordinate descent. 
	
	Recently, the application of DL to wireless communications problems has attracted significant attention \cite{pham2021intelligent, jagannath2021redefining, dai2020deep, zhu2020toward}, 
	with one of the considered problems being HBF design \cite{li2019deep, huang2019deep, hu2021two, elbir2019joint, elbir2019hybrid, peken2020deep, hojatian2021unsupervised, balevi2021unfolded, agiv2022learn, shi2022deep, chen2022hybrid}. Two typical DL techniques are often applied: purely data-driven DL and hybrid model-based DL~\cite{shlezinger2022model}. The former relies mainly on the learning capability of deep neural networks (DNNs) \cite{li2019deep, huang2019deep, hu2021two},  convolutional neural networks (CNNs) \cite{elbir2019joint, elbir2019hybrid, peken2020deep, hojatian2021unsupervised, elbir2021family, chen2020sub, hojatian2022flexible, elbir2021terahertz}, or deep reinforcement learning \cite{wang2020precodernet, hu2021joint} to generate HBF beamformers. For example, \cite{elbir2021terahertz} designed a mMIMO HBF with a group-of-subarray structure in the low-THz band via both model-based AO and data-driven CNNs. It was shown that while the former can achieve better performance, the latter operates approximately $500$ times faster than the model-based AO. Yet, such a purely data-driven DL approach has major limitations due to its resource constraints, high complexity, and black-box nature \cite{jagannath2021redefining,nguyen2020deep, nguyen2020application, nguyen2022leveraging, zappone2019wireless}. 
	
	Model-based DL encompasses a family of hybrid methodologies for combining domain knowledge with data to realize efficient inference mappings~\cite{shlezinger2022model2}. A leading hybrid methodology is {\em deep unfolding}, which leverages DL techniques to improve model-based iterative optimizers in terms of convergence, robustness, and performance~\cite{monga2021algorithm}. 
	In the context of HBF design, Balevi \textit{et al.} \cite{balevi2021unfolded} used deep generative unfolding models to obtain near-optimal hybrid beamformers with reduced feedback and complexity. Luo \textit{et al.} \cite{luo2022mdl} and Shi \textit{et al.} \cite{shi2022deep} proposed deep unfolding HBF solutions based on unfolding AO and iterative gradient descent, respectively. 
	
	Most of the aforementioned works focused on HBF design in conventional narrowband  systems. In wideband  MIMO-OFDM systems, the analog beamformer is typically frequency flat, i.e., a common analog beamforming matrix must serve the entire frequency band. This imposes extra difficulties on the HBF design, and the approaches proposed for narrowband systems are not readily applicable. 
	The work 
	\cite{agiv2022learn} proposed a low-complexity HBF design by unfolding the projected gradient ascent (PGA) optimization with a fixed number of iterations and learning the hyperparameters of the iterative optimizer from the data.
	Chen \textit{et al.} \cite{chen2022hybrid} proposed a DNN architecture that unfolds the weighted minimum mean square error manifold optimization using  fully-connected DNNs to learn the step size in each iteration, leading to faster convergence and improved performance. However, high complexity is still required to update the gradient and the solutions in each iteration.   
	Kang \textit{et al.} \cite{kang2022mixed} introduced a deep unfolding hybrid beamforming design induced by a stochastic successive convex approximation algorithm. This scheme achieves good HBF performance; however, its highly-parameterized DNN network architecture is complicated, and the use of black-box DNNs results in high complexity. In \cite{pang2022mggan}, a DNN model referred to as a multi-generator generative adversarial network (MGGAN) was introduced for HBF design with rank-deficient channels. Similar to \cite{kang2022mixed}, the MGGAN architecture is highly complex. 
	
	\subsection{Contributions}
	
	
	In this paper, we propose efficient deep unfolding approaches for the designs of both fully-connected HBF (FC-HBF) and dynamic sub-connected HBF (SC-HBF) architectures. The proposed deep unfolding frameworks are based on unrolling iterations of the MO-AltMin algorithm of
	\cite{yu2016alternating}, and they are thus referred to as \textit{ManNet}-based HBF. The main idea is to first transform the challenging SE maximization problem into an approximate matrix factorization problem, in which both the analog and digital precoders admit LS formulations. In each iteration, the analog beamformers are produced by a DNN, while the digital beamformers are obtained via closed-form LS solutions. Furthermore, the employed DNN has a low-complexity sparsely-connected structure based on unfolding the projected gradient descent (PGD) algorithm. In this sense, the proposed ManNet-based HBF designs are a two-fold deep unfolding procedure. We summarize our main contributions as follows:
	
	\begin{itemize}
		
		\item We propose an unfolding framework for the design of FC-HBF architectures based on unfolding MO-AltMin. Unlike most existing DL-aided FC-HBF designs, the unfolding framework is developed by investigating the matrix factorization problem for HBF design rather than the original SE maximization. Thereby the complicated log-det objective function is transformed into a simpler norm-squared form in which the digital and (vectorized) analog precoders are alternately solved via LS. This significantly simplifies the design and reduces the overall complexity compared to the unfolding methods in \cite{agiv2022learn, kang2022mixed}.
		
		\item Based on the unfolded framework, we develop a lightweight DNN architecture called ManNet to estimate the analog beamformer based on PGD. ManNet is a sparsely connected DNN with an explainable architecture and low-complexity operations. Specifically, it can output reliable analog precoding coefficients with only a few layers, each requiring only element-wise multiplications between the input and weight vectors. We also propose an efficient unsupervised training procedure for ManNet. The training strategy offers fast convergence with limited  training data and no training labels.
		
		\item We then focus on dynamic SC-HBF design. The trained ManNet can be readily applied here. Specifically, we propose a low-complexity scheme to establish the dynamic connections between the RF chains and antennas, and the sparse analog precoding matrix is obtained by matching the channel gains with the output of ManNet. To further reduce the complexity of the SC-HBF design, we develop a simplified version of ManNet, referred to as subManNet, to directly output the sparse analog precoder for SC-HBF. The proposed schemes can also be applied to the fixed SC-HBF architecture.
		
		\item We present simulation results demonstrating that the ManNet-based FC-HBF scheme attains better performance in much less time and with much lower computational complexity than the conventional MO-AltMin \cite{yu2016alternating} and AO \cite{sohrabi2016hybrid} approaches. In particular, the proposed ManNet and subManNet-aided SC-HBF schemes achieve performance similar to that of FC-HBF, and much better than semideﬁnite relaxation-based alternating minimization (SDR-AltMin) \cite{yu2016alternating}. 
	\end{itemize}
	
	\subsection{Paper Organization and Notation}
	The rest of the paper is organized as follows.  Section \ref{sec_system_model} presents the signal and channel models, and the considered design problems. Sections \ref{sec_FCHBF} and \ref{sec_SCHBF} detail the proposed FC-HBF and SC-HBF designs, respectively. Numerical results are given in Section \ref{sec_sim}, while Section \ref{sec_conclusion} concludes the paper. 
	
	Throughout the paper, numbers, vectors, and matrices are denoted by lower-case, boldface lower-case, and boldface upper-case letters, respectively, while $[\mA]_{i,j}$ represents the $(i,j)$-th entry of  matrix $\mA$. We denote by $(\cdot)^T$ and $(\cdot)^H$ the transpose and the conjugate transpose of a matrix or vector, respectively, and $\mA^{\dagger}$ is the pseudo-inverse of a matrix $\mA$. The matrix $\mathrm {diag} \{ \va_1, \ldots, \va_N \}$ is block diagonal with diagonal columns $\va_1, \ldots, \va_N$. Furthermore, $\abs{\cdot}$ denotes either the absolute value of a scalar or the cardinality of a set, and $\odot$ represents the Hadamard product. {$\mathcal{(C)N}(\mu, \sigma^2)$ denotes a (complex) normal distribution with mean $\mathbf{\mu}$ and variance $\sigma^2$, while $\mathcal{U}[a,b]$ denotes a uniform distribution over given range $[a,b]$.}
	
	\section{Signal Model and Problem Formulation}
	\label{sec_system_model}
	
	
	\subsection{Signal Model}
	We consider the downlink of a point-to-point wideband mMIMO-OFDM system, where the base station (BS) and the mobile station (MS) are equipped with $\Nt$ and $\Nr$ antennas, respectively. Let $\bs[k] \in {\mathbb{C}}^{\Ns \times 1}$ denote the $\Ns$-dimensional transmit vector from the BS to the MS on the $k$-th subcarrier, with $\mean{\bs[k] \bs[k]^H}=\bI_{\Ns}$, $k = 1,2,\ldots,K$, where $K$ is the number of subcarriers. The BS employs a frequency-flat analog precoder $\Frf \in \mathbb{C}^{\Nt \times \Nrf}$ and a frequency-dependent digital baseband precoder $\Fbb[k] \in \mathbb{C}^{\Nrf \times \Ns}$, where $\Nrf$ is the number of RF chains at the BS, $\Ns \leq \Nrf \leq \Nt$, and the normalized transmit power constraint at the BS is given as $\norm{\Frf \Fbb[k]}_F^2 = \Ns, \forall k$. To focus on the design of hybrid precoders, we assume that $\Nr$ is relatively small so that a fully digital combiner $\mV[k] \in {\mathbb{C}}^{\Nr \times \Ns}$ is employed at the MS receiver for the $k$-th subcarrier. The post-processed signal at the MS is expressed as 
	\begin{align}
		\vy[k] &= \sqrt{\rho} \mV[k] \mH[k] \Frf \Fbb[k] \bs[k] + \mV[k]^H \bn[k], \label{processed_received_signal}
	\end{align}
	where $\rho$ denotes the average received power, $\bn[k] \sim \mathcal{CN}(\mathbf{0},\sigma^2_{\text{n}} \mI_{\Nr})$ is additive white Gaussian noise (AWGN) at the MS, and $\mH[k]$ is the channel matrix at the $k$-th subcarrier. 
	
	We adopt the extended Saleh-Valenzuela channel model and express $\mH[k]$ as \cite{yu2016alternating} 
	\begin{align}
		\mH[k] =\xi   \sum_{p=1}^{P} \alpha_{p} e^{-j2\pi \tau_p f_k} \ar(\theta^\text{r}_{p}, \phi^\text{r}_{p}, f_k) \at(\theta^\text{t}_{p}, \phi^\text{t}_{p}, f_k)^H. \label{eq_channel_model}
	\end{align} 
	In \eqref{eq_channel_model}, $\xi = \sqrt{\frac{N_{\rm r}N_{\rm t}}{P}}$ and $f_k = f_{\text{c}} + \frac{\text{BW}(2k-1-K)}{2K}$ where $\text{BW}$ and $f_{\text{c}}$ represent the system bandwidth and center frequency; $P$ is the number of propagation paths; $\alpha_{p}$ and $\tau_{p}$ are the complex gain and time-of-arrival (ToA) of the $p$-th path; $\phi^\text{t}_{p} (\theta^\text{t}_{p})$ and $\phi^\text{r}_{p} (\theta^\text{r}_{p})$ represent the azimuth (elevation) angles of departure (AoDs) and arrivals (AOAs) of the $p$-th path; $\at \in \mathbb{C}^{\Nt \times 1}$ and $\ar \in \mathbb{C}^{\Nr \times 1}$ denote the transmit and receive array response vectors, respectively. We assume that the BS is equipped with a UPA of size  $\Nth \times \Ntv$, where $\Nth$ and $\Ntv$ are the numbers of antennas in the horizontal and vertical dimensions, and $\Nth \Ntv = \Nt$. We assume half-wavelength antenna spacing at the BS, and thus, $\at(\theta^\text{t}_{p},\phi^\text{t}_{p},f_k)$ is given as \cite{yu2016alternating}
	\begin{align}
		&\at(\theta^\text{t}_{p},\phi^\text{t}_{p},f_k) =\frac{1}{\sqrt{\Nr}}\Big[
		1,\dots,e^{j \pi \frac{f_k}{f_{\text{c}}} (i_{\text{h}} \sin(\phi^\text{t}_{p}) \sin(\theta^\text{t}_{p}) + i_{\text{v}} \cos(\theta^\text{t}_{p}) )},\notag\\
		&\qquad \dots,e^{j\pi \frac{f_k}{f_{\text{c}}} ((\Nrh-1) \sin(\phi^\text{t}_{p}) \sin(\theta^\text{t}_{p}) + (\Nrv-1) \cos(\theta^\text{t}_{p}))} \Big]^T, \label{eq_array_response}
	\end{align}
	where $i_{\text{h}} \in [0, \Nth)$ and $i_{\text{v}} \in [0, \Ntv)$ denote the antenna indices on the horizontal and vertical dimensions, respectively. The array response vector $\ar(\theta^\text{r}_{p},\phi^\text{r}_{p}, f_k)$ at the MS are modeled similarly.
	
	\subsection{FC-HBF and SC-HBF Architectures}
	\label{sec_beam_squint}
	
	We consider both FC-HBF and SC-HBF phase-shifter-based architectures. In the former, each RF chain is connected to all $\Nt$ antennas, requiring a total of $\Nrf \Nt$ phase shifters. In this case, the analog precoder is constrained as
	\begin{align*}
		\Frf \in \Afull \triangleq \left\{ \Frf : [\Frf]_{m, n} = e^{j \zeta_{m, n}},\  \forall m, n \right\}, \nbthis \label{eq_feasible_set_full}
	\end{align*}
	where $\zeta_{m, n}$ represents the effect of the phase shifter between the $n$-th RF chain and the $m$-th antenna. 
	
	In the SC-HBF architecture, each RF chain only connects to a subset of $M \triangleq \frac{\Nt}{\Nrf}$ antennas to reduce the hardware complexity and power consumption (assuming that $\frac{\Nt}{\Nrf}$ is an integer for simplicity). Such an analog network requires only $\Nt$ phase shifters in total, which is a factor of $\Nrf$ lower than FC-HBF. We assume a dynamic sub-connected architecture in which RF chains are connected to non-overlapping subsets of antennas. In this case, the sub-connected analog precoder is constrained as
	\begin{align*}
		&\Frf \in \Asub \triangleq \Big\{\Frf : [\Frf]_{m, n} \in  \left\{0, e^{j \zeta_{m, n}} \right\}, \\ 
		&\quad \sum_{m=1}^{\Nt} \abs{[\Frf]_{m, n}} = M,
		\sum_{n=1}^{\Nrf} \abs{[\Frf]_{m, n}} = 1, \ \forall m, n \Big\}, \nbthis \label{eq_feasible_set_sub}
	\end{align*}
	i.e., the $(m, n)$-th entry of $\Frf$ can be either a non-zero (unit-modulus) coefficient, when the $n$-th RF chain is connected to the $m$-th antenna, or zero otherwise. Furthermore, in each row and column of $\Frf$, there are only a single and $M$ nonzero elements, respectively. Note that the conventional fixed SC-HBF architecture is a special case of the dynamic one, i.e., when the $n$-th RF chain is connected to $M$ adjacent antennas indexed from $(n-1)M + 1$ to $n M$. In this case, we have $\Frf = \text{blkdiag} \left\{ \bar{\vf}_1, \ldots, \bar{\vf}_n, \ldots, \bar{\vf}_{\Nrf}  \right\}$, where $\bar{\vf}_n = \left[f_{1,n}, \ldots, f_{M, n} \right]^T$, as considered in \cite{yu2016alternating}. 
	
	Compared to the fixed SC-HBF architecture, the dynamic approach additionally requires $\Nt$ switches in the analog precoding network to dynamically configure the connections between the RF chains and the antennas. However, the switches do not significantly impact the total power consumption of the system. The power consumption of a typical switch is 6 times less than that of a phase shifter and $40$ times less than a digital-to-analog converter (DAC) \cite{nguyen2019unequally, mendez2016hybrid}. Furthermore, low-power, low-cost, and high-speed tunable switches can be used \cite{mendez2016hybrid, zhu2016adaptive, schmid2014analysis} in dynamic SC-HBF structures. 
	
	\subsection{Problem Formulation} 
	Based on \eqref{processed_received_signal}, the average per-subcarrier achievable SE for Gaussian symbols is given by \cite{yu2016alternating}
	\begin{align*}
		R &= \frac{1}{K} \sum_{k=1}^{K} \log_2 \text{det} \Big( \bI_{\Ns} + \frac{\rho}{\sigma^2_{\text{n}} \Ns} \bV[k]^{\dagger} \mH[k] \Frf \Fbb[k]  \\
		&\hspace{3cm} \times \Fbb[k]^H \Frf^H \mH[k]^H \bV[k] \Big).  \nbthis \label{eq_SE}
	\end{align*}
	We aim at designing the precoders and combiners $\{\Frf, \Fbb[k], \mV[k] \}$ to maximize $R$, which is challenging due to the strong coupling among the variables. However, given $\{\Frf, \Fbb[k]\}$, the optimal solution for $\mV[k]$ is the matrix whose columns are the $\Ns$ left singular vectors corresponding to the $\Ns$ largest singular values of $\mH[k] \Frf \Fbb[k]$ \cite{tse2005fundamentals}. Therefore, we focus on the designs of the hybrid precoders $\{\Frf, \Fbb[k]\}$ in the sequel. 
	
	The SE maximizing hybrid precoding design can be approximately achieved via the following optimization \cite{el2014spatially, yu2016alternating}:
	\begin{subequations}
		\label{opt_prob}
		\begin{align*} 
			\quad \underset{\substack{ \Frf, \{\Fbb[k]\}_{k=1}^K }}{\textrm{minimize}} \quad &  \sum_{k=1}^{K}\norm{\bF_{\text{opt}}[k] - \Frf \Fbb[k]}_{\mathcal{F}} \nbthis \label{obj_func} \\
			\textrm{subject to} \quad
			&\Frf \in \mathcal{A}, \nbthis \label{cons_3}\\
			&\norm{\Frf \Fbb[k]}_\mathcal{F}^2 = \Ns,\ \forall k, \nbthis \label{cons_2}
		\end{align*}
	\end{subequations}
	where $\bF_{\text{opt}}[k] \in \setC^{\Nt \times \Ns}$ is the unconstrained optimal digital precoder at the $k$-th subcarrier, whose columns are the $\Ns$ right singular vectors corresponding to the $\Ns$ largest singular values of $\bH[k]$ and scaled with water-filling power factors. In \eqref{cons_3}, the feasible set $\mathcal{A}$ of the analog precoder  can be either $\Afull$ or $\Asub$, defined in \eqref{eq_feasible_set_full} and \eqref{eq_feasible_set_sub}, respectively, depending on the HBF architecture. This constraint enforces the unit modulus of the analog precoding coefficients and the configuration of the sub-connected analog network. The per-subcarrier transmit power is constrained in \eqref{cons_2}. 
	
	Problem \eqref{opt_prob} is a non-convex matrix factorization problem, and joint optimization of $\Frf$ and $\{\Fbb[k]\}_{k=1}^K$ is complicated due to constraint \eqref{cons_3}. MO-AltMin \cite{yu2016alternating} and OMP \cite{el2014spatially} are two conventional model-based algorithms for tackling \eqref{opt_prob}. As discussed earlier, MO-AltMin is highly complex and converges slowly when the system dimensions are large. In contrast, OMP maintains low complexity, but it has unsatisfactory performance. We overcome these deficiencies by proposing an efficient deep unfolding approach next.
	
	\section{Proposed FC-HBF Design}
	\label{sec_FCHBF}
	We first focus on the design of FC-HBF, i.e., the design in \eqref{opt_prob} with $\Frf \in \Afull$. To this end, we propose a deep unfolding approach referred to as ManNet-based FC-HBF. Its main idea is to unfold the MO-AltMin algorithm, estimating the solution to $\Frf$ using ManNet, an unfolding DNN designed based on PGD optimization. 
	
	\subsection{Proposed ManNet-Based FC-HBF Approach}
	
	\subsubsection{Main Idea}
	
	In the proposed approach, we apply the iterative alternating minimization method of \cite{yu2016alternating}. Specifically, in each iteration, we first optimize $\Frf$ with $\Fbb[k]$ given and constraint \eqref{cons_2} omitted. Then we design $\Fbb[k]$ to meet the constraint given the optimized $\Frf$. Thus, we first consider the following problem:
	\begin{subequations}
		\label{opt_prob_RF}
		\begin{align} 
			\quad \underset{\substack{ \Frf }}{\textrm{minimize}}\ & \sum_{k=1}^{K}\norm{\bF_{\text{opt}}[k] - \Frf \Fbb[k]}_{\mathcal{F}}^2,\\
			\textrm{subject to}\ &\Frf \in \Afull, \nbthis \label{cons_full}
		\end{align}
	\end{subequations}
	where the quadratic form of the objective function is introduced without affecting the solution. Let us denote
	\begin{align*}
		\tilde{\vx} &\triangleq \text{vec}(\Frf) \in \setC^{\Nt\Nrf \times 1}, \nbthis \label{eq_trans_x} \\
		\tilde{\vz}[k] &\triangleq \text{vec}(\bF_{\text{opt}}[k]) \in \setC^{\Nt\Ns \times 1}, \nbthis \label{eq_trans_z}\\
		\tilde{\bB}[k] &\triangleq (\Fbb[k])^T \otimes \bI_{\Nt}  \in \setC^{\Nt\Ns \times \Nt\Nrf}. \nbthis \label{eq_trans_B}
	\end{align*}
	Then, the objective function in \eqref{opt_prob_RF} can be re-expressed as
	\begin{align*}
		\sum_{k=1}^{K} \norm{\bF_{\text{opt}}[k] - \Frf \Fbb[k]}_{\mathcal{F}}^2 = \sum_{k=1}^{K} \normshort{\tilde{\vz}[k] - \tilde{\bB}[k] \tilde{\vx}}^2. \nbthis \label{eq_trans_obj}
	\end{align*}
	Furthermore, by denoting
	\begin{align*}
		\vx &\triangleq  \begin{bmatrix}
			\re{\tilde{\vx}} \\
			\im{\tilde{\vx}}
		\end{bmatrix}  \in \setR^{2\Nt\Nrf \times 1},  \nbthis \label{eq_def_x}\\
		\vz[k] &\triangleq  \begin{bmatrix}
			\re{\tilde{\vz}[k]} \\
			\im{\tilde{\vz}[k]}
		\end{bmatrix} \in \setR^{2\Nt\Ns \times 1},  \nbthis \label{eq_def_z} \\
		\bB[k] &\triangleq \begin{bmatrix}
			\re{\tilde{\bB}[k]} & -\im{\tilde{\bB}[k]}\\
			\im{\tilde{\bB}[k]} & \re{\tilde{\bB}[k]}
		\end{bmatrix} \in \setR^{2\Nt\Ns \times 2\Nt\Nrf}, \nbthis \label{def_B}
	\end{align*}
	with $\re{\cdot}$ and $\im{\cdot}$ representing the real and imaginary parts of a complex vector/matrix, respectively, we can write
	\begin{align*}
		\sum_{k=1}^{K}\norm{\bF_{\text{opt}}[k] - \Frf \Fbb[k]}_{\mathcal{F}}^2 = \sum_{k=1}^{K} \norm{{\vz[k]} - {\bB[k]} {\vx}}^2. \nbthis \label{eq_trans_obj_real}
	\end{align*}
	
	Define the transformation
	\begin{align*}
		\mathcal{V} : \Frf \rightarrow \vx \text{~and~} \mathcal{V}^{-1} : \vx \rightarrow \Frf \nbthis \label{def_transformation}
	\end{align*}
	which transforms the complex-valued matrix $\Frf$ into the real-valued vector $\vx$ and vice versa, respectively. With the newly introduced variables, the optimal solution to problem \eqref{opt_prob_RF} admits the LS form 
	\begin{align*}
		\vx^{\star} = \argmin_{\vx: \mathcal{V}^{-1} (\vx) \in \Afull} \sum_{k=1}^{K} \norm{\vz[k] - \bB[k] \vx}^2. \nbthis \label{eq_opt_sol}
	\end{align*}
	
	Based on \eqref{eq_opt_sol}, a deep unfolding DNN of $L$ layers is designed to mimic the PGD algorithm to approximate $\vx^{\star}$. Specifically, let $\bx_{\ell}$ be the output of the $\ell$-th layer of the DNN. From \eqref{eq_opt_sol}, $\bx_{\ell}$ can be produced as \cite{samuel2019learning}
	\begin{align*}
		\bx_{\ell} &= \mathcal{T}_{\ell} \left( \bx - \delta_{\ell} \frac{\partial \sum_{k=1}^{K} \norm{{\vz[k]} - {\bB[k]} {\vx}}^2}{\partial \vx} \right)_{\vx = \vx_{\ell - 1}} \\
		&= \mathcal{T}_{\ell} \left( \vx_{\ell - 1} - \sum_{k=1}^{K} \left(\delta_{\ell} \bB[k]^T \vz[k] + \delta_{\ell} \bB[k]^T \bB[k] \vx_{\ell - 1}\right) \right)\\
		&= \mathcal{T}_{\ell} \left( \vx_{\ell - 1} - \delta_{\ell} \bar{\vz} + \delta_{\ell} \sum_{k=1}^{K} \bar{\mB}[k] \vx_{\ell - 1} \right), \nbthis \label{eq_unfolding}
	\end{align*} 
	where $\delta_{\ell}$ denotes a step size, $\mathcal{T}_{\ell} (\cdot)$ represents a nonlinear projection operator, and in the last equality we denote $\bar{\vz} \triangleq \sum_{k=1}^{K} \bB[k]^T \vz[k]$ and $\bar{\mB}[k] \triangleq \bB[k]^T \bB[k], \forall k$. The relationship in \eqref{eq_unfolding} motivates a DNN model to learn $\vx^{\star}$ wherein the output of a given layer (i.e., $\bx_{\ell}$ in the $\ell$-th layer) results from a nonlinear projection applied to the output of the previous layer (i.e., $\bx_{\ell-1}$ in the $(\ell-1)$-th layer) and other given information, including $\bar{\vz}$ and $\{\bar{\bB}[k]\}$ which is short for $\{\bar{\bB}[k]\}_{k=1}^{K}$. The nonlinear projection is performed with trainable parameters, i.e., the weights of the DNN. Applied over multiple layers, the DNN can be structured and trained such that its final output, i.e., $\bx_{L}$, will be a good estimate of $\vx^{\star}$. 
	In the following, we develop such an efficient DNN architecture referred to as ManNet.
	
	\subsubsection{ManNet Architecture}
	
	\begin{figure}[t]
		\centering
		\includegraphics[scale=0.65]{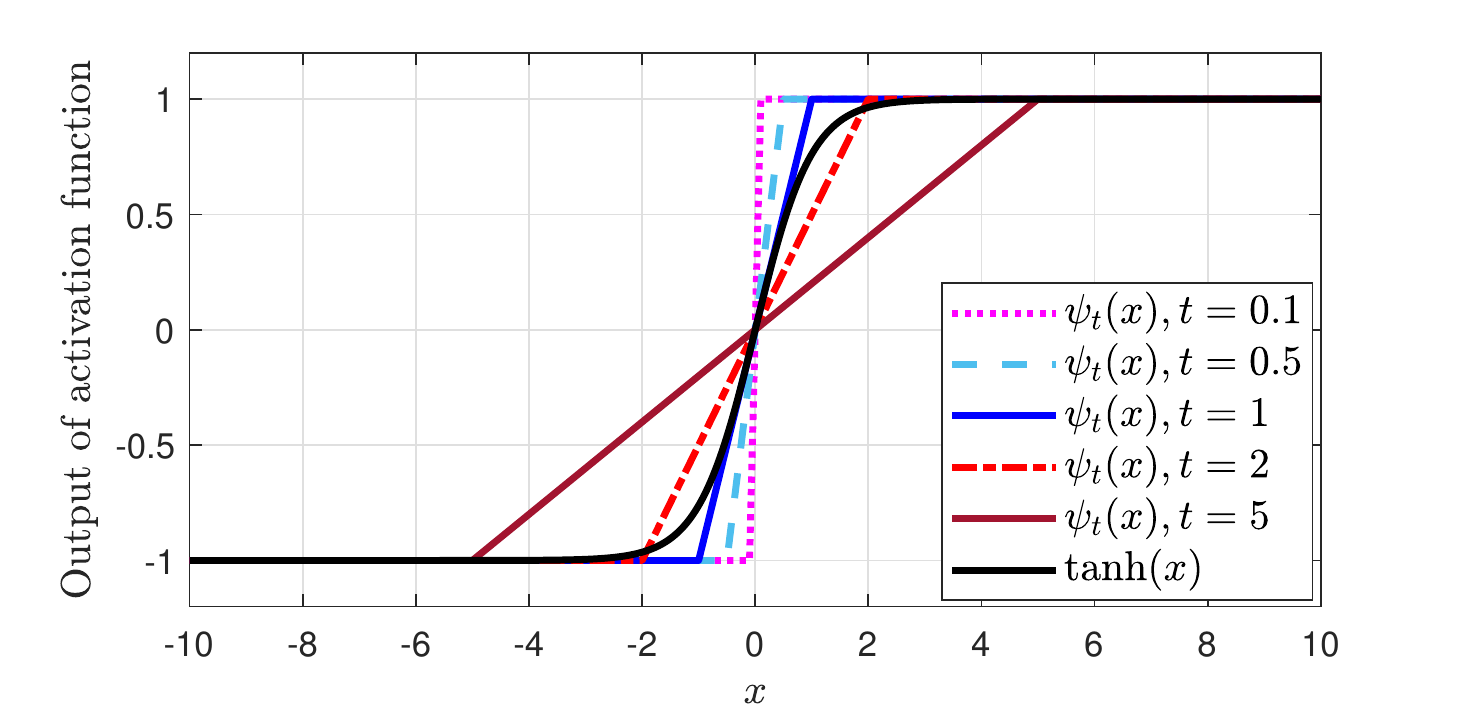}
		\caption{Activation functions $\psi_t(x)$ and $\mathrm{tanh}(x)$.}
		\label{fig_activation}
	\end{figure}
	
	Denote
	\begin{align*}
		\vu_{\ell-1} \triangleq - \bar{\vz} + \sum_{k=1}^{K} \bar{\bB}[k] \vx_{\ell - 1}, \nbthis \label{def_u}
	\end{align*}
	and rewrite \eqref{eq_unfolding} as
	\begin{align*}
		\bx_{\ell}
		= \mathcal{T}_{\ell} \left( \vx_{\ell - 1} + \delta_{\ell} \vu_{\ell-1}  \right). \nbthis \label{eq_unfolding_1}
	\end{align*}  
	We propose ManNet as a network of $L$ layers defined by \eqref{eq_unfolding_1} with the objective of learning $\vx^{\star}$. It takes $\vx_{\ell - 1}$ and $\vu_{\ell-1}$ as the input of the $\ell$-th layer, and outputs $\bx_{\ell}$ as the sum of the outputs of two other sub-networks based on the two input vectors $\vx_{\ell - 1}$ and $\vu_{\ell-1}$ in \eqref{eq_unfolding_1}. Importantly, the $n$-th element of $\bx_{\ell}$ only depends on the $n$-th elements of $\bx_{\ell-1}$ and $\vu_{\ell-1}$. Thus, only the nodes (or neurons) at the same vertical level between the layers are connected making ManNet a sparsely connected DNN. Furthermore, we define the activation function
	\begin{align*}
		\psi_t(x) = -1 + \frac{1}{\abs{t}} \left( \sigma(x+t) - \sigma(x-t) \right), \nbthis \label{def_activation}
	\end{align*}
	where $\sigma(\cdot)$ is the rectified linear unit (ReLU) activation function, and $t$ is a hyperparameter. This guarantees that the amplitudes of the elements of $\bx_{\ell}$ are in the range $[-1,1], \forall t$, i.e., $\abs{x_i} \leq 1, i = 1, \ldots, 2\Nt\Nrf$.\footnote{The activation function $\mathrm{tanh}(x) = \frac{e^{x} - e^{-x}}{e^{x} + e^{-x}}$ can also output values in $[-1,1]$, as seen in Fig.\ \ref{fig_activation}. However, its slope is fixed, causing a fixed mapping when applying the activation function. We have found via simulation that by proper fine-tuning of $t$, $\psi_t(x)$ provides better performance than $\mathrm{tanh}(x)$.} As a result, its corresponding complex-valued matrix representation, denoted as $\Frf^{(\ell)} = \mathcal{V}^{-1} (\bx_{\ell})$, has elements satisfying $\absshort{[\Frf^{(\ell)}]_{m,n}} \leq \sqrt{2},\ \forall m, n, \ell$. 
	As this does not immediately ensure $\Frf^{(\ell)} \in \Afull$ as constrained in \eqref{cons_full}, the final output of the DNN ($\bx_{L}$) is  normalized to produce a  solution $\Frf = \mathcal{V}^{-1} (\bx_{L})$ satisfying~\eqref{cons_full}.
	
	\begin{figure*}[t]
		\centering
		\includegraphics[scale=0.5]{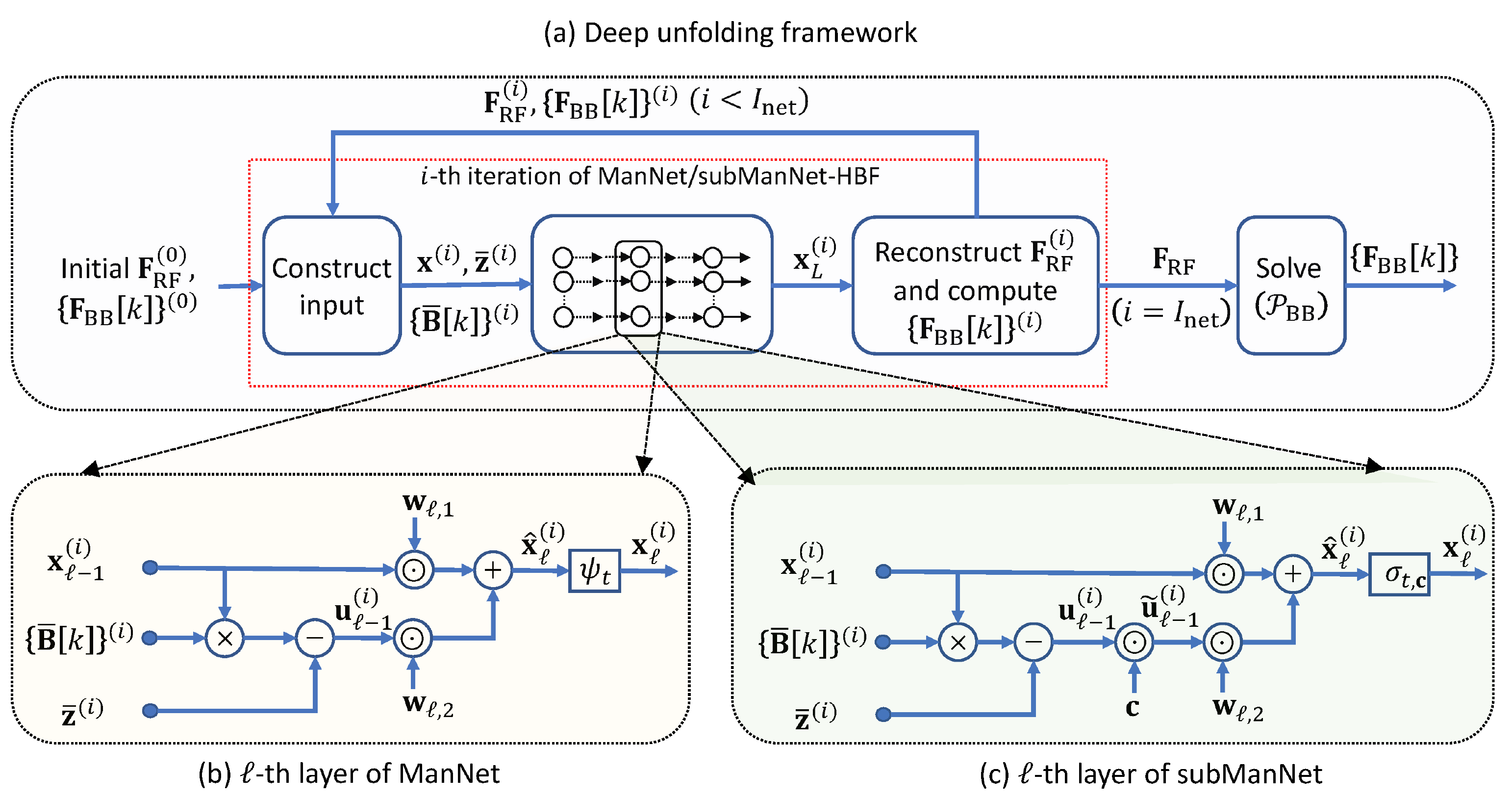}
		\caption{Illustration of (a) the proposed deep unfolding framework for FC-HBF and SC-HBF designs, the $\ell$-th layer of (b) ManNet and (c) subManNet.}
		\label{fig_iterative_unfolding}
	\end{figure*}
	
	Let $\{\vw_{\ell,1}, \vw_{\ell,2}\}_{\ell=1}^L$ be the weight vectors of the two sub-networks associated with inputs $\vx_{\ell - 1}$ and $\vu_{\ell-1}$ in the $\ell$-th layer of ManNet. A detailed network architecture illustrating the operation of each layer of ManNet is shown in Fig.\ \ref{fig_iterative_unfolding}(b). 
	
	\subsubsection{Training ManNet}
	
	We employ an unsupervised training approach for ManNet with the loss function
	\begin{align*}
		\hspace{-0.24cm}\mathcal{L} \left(\{\vw_{\ell,1}, \vw_{\ell,2}\}_{\ell=1}^L\right) = \sum_{\ell=1}^{L} \log(\ell) 
		\left(\sum_{k=1}^{K} \norm{\vz[k] - \bB[k] \vx_{\ell}}^2 \right), \nbthis \label{eq_loss}
	\end{align*}
	which sums the total weighted objective values of all $L$ layers. The DNN is trained to optimize the parameter set $\{\vw_{\ell,1}, \vw_{\ell,2}\}_{\ell=1}^L$ such that $\mathcal{L}\left(\{\vw_{\ell,1}, \vw_{\ell,2}\}_{\ell=1}^L\right)$ is minimized, which also directly minimizes the objective function in \eqref{eq_opt_sol} at the network output $\vx_{\ell} = \vx_L$. We note here that, otherwise, if supervised training were used, it would require the implementation of a conventional high-complexity HBF scheme to obtain the training labels, i.e., the analog precoding coefficients. This would dramatically increase the training complexity. Because optimal solutions to obtain the labels are unavailable, employing sub-optimal solutions for supervised training may limit the performance of ManNet.
	
	In Algorithm \ref{alg_train_ManNet}, we summarize the ManNet training process using a training data set $\mathcal{D}$. To initialize the training, the weight vectors are first randomly generated from the distribution $\mathcal{N}(0,0.01)$, and an initial learning rate is set. Then, ManNet is trained over $\mathcal{E}$ epochs, each using $\mathcal{B}$ batches $\{\mathcal{H}^{(b)}\}_{b=1}^{\mathcal{B}}$, where $\mathcal{H}^{(b)} = \left\{ \{\mH[k]\}_1, \ldots, \{\mH[k]\}_{\abs{\mathcal{H}^{(b)}}} \right\}$, and $\abs{\mathcal{H}^{(b)}}$ denotes the training batch size.  For the $b$-th batch, we randomly generate $\Frf^{(b,0)}$, and $\{\Fbb[k]\}^{(b,0)}$ is obtained via the LS solution
	\begin{align*}
		\Fbb[k]^{(b,i)} = (\Frf^{(b,i)})^{\dagger} \Fopt[k]^{(b)}, \forall k, b, i, \nbthis \label{eq_LS_solution}
	\end{align*}
	where $\Fopt[k]^{(b)}$ is the optimal fully digital precoder for the channels at the $k$-th subcarrier in $\mathcal{H}^{(b)}$, and $\mathbf{X}^{(b,i)}$ denotes the data $\mathbf{X}$ in the $b$-th batch of the $i$-th iteration.
	From step 6, the iterative process of optimizing the ManNet weights is performed. Specifically, in the $i$-th iteration, for given $\Frf^{(b,i)}$ and $\{\Fbb[k]\}^{(b,i)}$, the real-valued $\vx^{(b,i)}$, $\{\vz[k]^{(b,i)}\}$, and $\{\mB[k]^{(b,i)}\}$ are constructed based on \eqref{eq_trans_x}--\eqref{def_B} in step 7, allowing computation of $\bar{\vz}^{(b,i)}$ and $\{\bar{\mB}[k]^{(b,i)}\}$ in steps 8 and 9, respectively. Steps 10--16 update $\hat{\bx}_{\ell}^{(b,i)}$ and the loss value, which is then used in an optimizer to update the weights in step 18. It is seen that the training for each data batch is an iterative process over $\mathcal{I}_{\text{net}}^{\text{train}}$ iterations. After each iteration, $\Frf^{(b,i)}$ and $\{\Fbb[k]\}^{(b,i)}$ are updated and utilized for the next set of training iterations until $\mathcal{I}_{\text{net}}^{\text{train}}$ iterations are completed. This iterative approach is efficient in reducing the amount of training data and accelerating the convergence, as we empirically show in Section \ref{sec_sim}.

	\begin{algorithm}[t]
		\small
		\caption{Unsupervised Training in ManNet}
		\label{alg_train_ManNet}
		\begin{algorithmic}[1]
			\REQUIRE Training set $\mathcal{D}$ of channels.
			\ENSURE Network parameters $\{\vw_{\ell,1}, \vw_{\ell,2}\}_{\ell=1}^{L}$.
			\STATE Initialize weights $\{\vw_{\ell,1}^{(1,1)}, \vw_{\ell,2}^{(1,1)}\}_{\ell=1}^{L}$ and learning rate.
			\FOR{{$e = 1 \rightarrow \mathcal{E}$}}
			\STATE {Randomly divide $\mathcal{D}$ into $\mathcal{B}$ batches $\{\mathcal{H}^{(b)}\}_{b=1}^{\mathcal{B}}$.}
			\FOR{$b = 1 \rightarrow \mathcal{B}$}
			\STATE Obtain $\Fopt^{(b)}$, randomly initialize $\Frf^{(b,0)}$, and compute $\{\Fbb[k]^{(b,0)}\}$ based on \eqref{eq_LS_solution}. 
			\FOR{$i = 1 \rightarrow \mathcal{I}_{\text{net}}^{\text{train}}$}
			\STATE Obtain $\vx^{(b,i)}$, $\{\vz[k]^{(b,i)}\}$, and $\{\mB[k]^{(b,i)}\}$ from $\Fopt^{(b)}$, $\Frf^{(b,i-1)}$, and $\{\Fbb[k]^{(b,i-1)}\}$ based on \eqref{eq_trans_x}--\eqref{def_B}.
			\STATE Compute $\bar{\vz}^{(b,i)} = \sum_{k=1}^{K} (\bB[k]^{(b,i)})^T \vz[k]^{(b,i)}$.
			\STATE Compute $\bar{\mB}[k]^{(b,i)} \triangleq (\bB[k]^{(b,i)})^T \bB[k]^{(b,i)}, \forall k$.
			\STATE $\mathcal{L}^{(b,i)} = 0, \vx_{0}^{(b,i)}=\mathbf{0}$.
			\FOR{$\ell = 1 \rightarrow L$}
			\STATE $\vu_{\ell-1}^{(b,i)} = - \bar{\vz}^{(b,i)} + \sum_{k=1}^{K} \bar{\bB}[k]^{(b,i)} \vx_{\ell - 1}^{(b,i)}$.
			\STATE $\hat{\bx}_{\ell}^{(b,i)} = \vw_{\ell,1}^{(b,i)} \odot \vx_{\ell-1}^{(b,i)} + \vw_{\ell,2}^{(b,i)} \odot \vu_{\ell - 1}^{(b,i)}$.
			\STATE $\bx_{\ell}^{(b,i)} = \psi_t(\hat{\bx}_{\ell}^{(b,i)})$.
			\STATE Accumulate the average loss value of the batch over ManNet's layers based on \eqref{eq_loss}: $\mathcal{L}^{(b,i)} = \mathcal{L}^{(b,i)} + \log(\ell) \frac{1}{K \abs{\mathcal{H}^{(b)}}} \sum_{k=1}^{K}  \norm{\vz[k]^{(b,i)} - \bB[k]^{(b,i)} \vx_{\ell}^{(b,i)}}^2$.
			\ENDFOR
			\STATE $\mathcal{L}(\{\vw_{\ell,1}^{(b,i)}, \vw_{\ell,2}^{(b,i)}\}_{\ell=1}^{L}) = \mathcal{L}^{(b,i)}$.
			\STATE Obtain $\{\vw_{\ell,1}^{(b,i+1)}, \vw_{\ell,2}^{(b,i+1)}\}$ with an optimizer.
			\STATE Update $\Frf^{(b,i)} = \mathcal{V}^{-1}(\vx_{\ell}^{(b,i)})$ and compute $\Fbb[k]^{(b,i)}$ based on \eqref{eq_LS_solution}.
			\ENDFOR
			\ENDFOR
			\ENDFOR
			\STATE Return $\{\vw_{\ell,1}, \vw_{\ell,2}\} = \left\{\vw_{\ell,1}^{(\mathcal{B},\mathcal{I}_{\text{net}}^{\text{train}})}, \vw_{\ell,2}^{(\mathcal{B},\mathcal{I}_{\text{net}}^{\text{train}})}\right\}$
		\end{algorithmic}
	\end{algorithm}
	
	\subsection{Overall ManNet-Based FC-HBF Algorithm}
	
	\begin{algorithm}[t]
		\small
		\caption{ManNet-based FC-HBF}
		\label{alg_ManNet_HBF}
		\begin{algorithmic}[1]
			\REQUIRE $\mH, \Fopt$, ManNet's trained parameters $\left\{ \{\vw_{\ell,1}, \vw_{\ell,2}\}_{\ell=1}^L \right \}$.
			\ENSURE $\Frf, \{\Fbb[k]\}$.
			\STATE Initialize $\Frf^{(0)}$ and compute $\{\Fbb[k]^{(0)}\}$ based on \eqref{eq_LS_solution_test}.
			\FOR{$i = 1, \ldots, \mathcal{I}_{\text{net}}$}
			\STATE Obtain $\vx^{(i)}$, $\{\vz[k]^{(i)}\}$, and $\{\mB[k]^{(i)}\}$ from $\Frf^{(i-1)}$ and $\{\Fbb[k]^{(i-1)}\}$ based on \eqref{eq_trans_x}--\eqref{def_B}. Set $\vx_{0}^{(i)} = \mathbf{0}$.
			\STATE Compute $\bar{\vz}^{(i)} = \sum_{k=1}^{K} (\bB[k]^{(i)})^T \vz[k]^{(i)}$.
			\STATE Compute $\bar{\bB}[k]^{(i)} = (\bB[k]^{(i)})^T \bB[k]^{(i)}, \forall k$.
			\FOR{$\ell = 1 \rightarrow L$}
			\STATE Construct the input: $\vu_{\ell-1}^{(i)} = \sum_{k=1}^{K}  \bar{\bB}[k]^{(i)} \vx_{\ell - 1}^{(i)} - \bar{\vz}^{(i)}$.
			\STATE Apply weights: $\hat{\bx}_{\ell}^{(i)} = \vw_{\ell,1} \odot \vx_{\ell-1}^{(i)} + \vw_{\ell,2} \odot \vu_{\ell-1}^{(i)}$.
			\STATE Apply the activation function: $\bx_{\ell}^{(i)} = \psi_t(\hat{\bx}_{\ell}^{(i)})$.
			\ENDFOR
			\STATE Reconstruct the complex RF precoding matrix $\Frf^{(i)}$ from $\bx_{L}^{(i)}$, i.e., $\Frf^{(i)} = \mathcal{V}^{-1}(\vx^{(i)})$.
			\STATE For iterations $1, \ldots, \mathcal{I}_{\text{net}} - 1$, compute $\{\Fbb[k]^{(i)}\}$ based on \eqref{eq_LS_solution_test}. For the last iteration, i.e., $i = \mathcal{I}_{\text{net}}$, set $\Frf = \Frf^{(\mathcal{I}_{\text{net}})}$ and obtain $\{\Fbb[k]\}$ based on \eqref{eq_Fbb}.
			\ENDFOR
		\end{algorithmic}
	\end{algorithm}
	
	Once the offline training process is completed, ManNet with the trained weight vectors is readily applied to online FC-HBF design. We refer to this approach as ManNet-based FC-HBF, and it is summarized in Algorithm \ref{alg_ManNet_HBF}. Specifically, we generate the initial analog precoder and compute the digital one in step 1. From step 2, the unfolding HBF design is performed over $\mathcal{I}_{\text{net}}$ iterations. In steps 3--5, $\vx$, $\{\vz[k]\}$, and $\{\mB[k]\}$ are obtained to compute $\bar{\vz}$ and $\{\bar{\mB}[k]\}$ in steps 4 and 5, respectively. After that, ManNet iteratively executes steps 6--10 to construct the outputs of its layers. Note that only element-wise multiplications between the weight and input vectors are required, as seen in step 8 and Fig.\ \ref{fig_iterative_unfolding}. The final output of ManNet, i.e., $\bx_{L}$, is reconstructed as the feasible solution to $\Frf$ in step 11, and the $\Fbb[k]$ are updated via LS, i.e.,
	\begin{align*}
		\Fbb[k]^{(i)} = (\Frf^{(i)})^{\dagger} \Fopt[k],\ \forall k, i. \nbthis \label{eq_LS_solution_test}
	\end{align*}
	The solutions for $\Frf$ and $\Fbb[k]$ are then utilized for the next iteration until $\mathcal{I}_{\text{net}}$ iterations are completed. Finally, with $\Frf$ obtained, the optimal digital precoder directly maximizing the SE in \eqref{eq_SE} can be solved by the problem
	\begin{subequations}
		\label{opt_prob_BB}
		\begin{align*} 
			\left(\mathcal{P}_{\text{BB}}\right):\ \underset{\substack{ \{\Fbb[k]\} }}{\textrm{maximize}} \quad &  R_{\text{BB}} \left(\{\Fbb[k]\}\right) \nbthis \label{obj_func_BB} \\
			\textrm{subject to} \quad
			&\tr{\mQ \Fbb[k] \Fbb[k]^H} = \Ns,\ \forall k, \nbthis \label{cons_2_BB}
		\end{align*}
	\end{subequations}
	where 
	\begin{align*}
		R_{\text{BB}} & \left(\{\Fbb[k]\}\right) \triangleq  \\  & \frac{1}{K}\sum_{k=1}^{K}\log_2 \det \left(\bI_{\Ns} + \frac{\rho}{\sigma^2_{\text{n}} \Ns} \tilde{\mH} \Fbb[k]  \Fbb[k]^H \tilde{\mH}^H\right),
	\end{align*}
	$\tilde{\mH} \triangleq \mH \Frf$, and $\mQ \triangleq \Frf^H \Frf$. This problem  has a well-known water-filling solution:
	\begin{align*}
		\Fbb[k] = \mQ^{-\frac{1}{2}} \tilde{\mU} \tilde{\boldsymbol{\Gamma}}, \nbthis \label{eq_Fbb}
	\end{align*}
	where the columns of $\tilde{\mU}$ are taken from the right singular vectors corresponding to the $\Ns$ largest singular values of $\tilde{\mH} \mQ^{-\frac{1}{2}}$, and $\tilde{\boldsymbol{\Gamma}}$ is a diagonal matrix whose elements are defined by the power allocated to the $\Ns$ data streams \cite{sohrabi2016hybrid}. In Algorithm \ref{alg_ManNet_HBF}, the final solution to $\{\Fbb[k]\}$ is obtained based on \eqref{eq_Fbb} in the last iteration, as shown in step 12. We illustrate the entire proposed deep unfolding framework of the ManNet-based FC-HBF design in Fig.\ \ref{fig_iterative_unfolding}(a).
	
	We note that the modular architecture of our unfolded network allows numbers of iterations in the training and online application phases of ManNet, i.e., $\mathcal{I}_{\text{net}}^{\text{train}}$ and $\mathcal{I}_{\text{net}}$ in Algorithms \ref{alg_train_ManNet} and \ref{alg_ManNet_HBF}, respectively, to be different. 
	In particular, we noted that during training, where the goal is to set the weights of ManNet, reliable learning can be achieved with just a few iterations, e.g., $\mathcal{I}_{\text{net}}^{\text{train}} = 3$, which are also enough for fast convergence. 
	During inference, when the goal is to set the hybrid precoders, the setting of $\mathcal{I}_{\text{net}}$ can balance performance-complexity tradeoff: while  the performance of the ManNet-based FC-HBF scheme improves with $\mathcal{I}_{\text{net}}$, its computational complexity linearly increases with $\mathcal{I}_{\text{net}}$, as will be shown next.

	\subsection{Complexity Analysis}
	\label{sec_comp_analysis_ManNet_HBF}
	We herein analyze the computational complexity of the proposed ManNet-based FC-HBF scheme in Algorithm \ref{alg_ManNet_HBF}. It is observed from \eqref{eq_trans_B} and \eqref{def_B} that $\mB[k]$ is a sparse matrix, in which only $2\Nrf$ and $2\Ns$ (out of $2\Nt\Nrf$ and $2\Nt\Ns$) elements in each row and column, respectively, are nonzero real-valued numbers. Thus, the complexity for computing $\bar{\vz}$ and $\{\bar{\mB}[k]\}$ in steps 4 and 5 is only $\mathcal{O}(K\Ns\Nrf)$ and $\mathcal{O}(K \Nrf^2 \Ns)$, respectively. Furthermore, $\bar{\mB}[k]$ has only $2\Nrf$ nonzero elements in each row and column, and hence step 7 requires a complexity of $\mathcal{O}(\Nt + 2K\Ns\Nrf)$. The weighting in step 8 performs only element-wise vector multiplication/addition, which has a complexity of $3\mathcal{O}(\Nt\Nrf)$. In step 12, obtaining $\{\Fbb[k]\}$ with \eqref{eq_LS_solution_test} has a complexity of $\mathcal{O}(\Nt K \Nrf^2 )$, while the complexity of \eqref{eq_Fbb} is $2 \mathcal{O} (\Nt  K \Nrf )$. As a result, the total complexity of Algorithm \ref{alg_ManNet_HBF} can be approximated as
	\begin{align*}
		&\mathcal{C}_{\textrm{ManNet-FCHBF}} = \left(\mathcal{I}_{\text{net}} - 1\right) \mathcal{O} (\Nt K  \Nrf^2 ) + \mathcal{O} (\Nt  K \Nrf )  \\ 
		&\qquad + \mathcal{I}_{\text{net}} \mathcal{O} (2 K \Nrf^2 \Ns + L (3\Nt\Nrf + 2K \Nrf \Ns) ). \nbthis \label{eq_comp_ManNetHBF}
	\end{align*}

	Compared to MO-AltMin \cite{yu2016alternating}, AO \cite{sohrabi2016hybrid, nguyen2020deep}, and OMP \cite{el2014spatially}, the proposed ManNet-based FC-HBF scheme has low complexity.  These approaches require complexities of 
	\begin{align*}
		&\mathcal{C}_{\textrm{MO-AltMin}} = \\ & \quad \mathcal{I}_{\text{MO}}^{\text{out}}  \mathcal{O} \left(\Nt K \Nrf^2  +  \mathcal{I}_{\text{MO}}^{\text{in}} (3\Nt\Nrf + 2 K (\Nrf^2 + \Nrf) \Ns)\right), \\
		&\mathcal{C}_{\text{AO}} = 2 \mathcal{O} (\Nt  K \Nrf ) + \mathcal{I}_{\text{AO}} \mathcal{O}(2\Nt^2 \Nrf^2), \\
		&\mathcal{C}_{\text{OMP}} = \mathcal{O}(\Nt K \Nrf^2  + 2 \Nt P \Ns + 4 \Nt \Nrf^2 + 4\Nt\Nrf\Ns)
	\end{align*}
	respectively, where $\mathcal{I}_{\text{MO}}^{\text{in}}$, $\mathcal{I}_{\text{MO}}^{\text{out}}$, and $\mathcal{I}_{\text{AO}}$ denote the number of inner and outer iterations for MO-AltMin and the number of iterations for AO, respectively. The number of iterations for the analog precoding designs in these schemes is $\mathcal{I}_{\text{MO}}^{\text{out}} \mathcal{I}_{\text{MO}}^{\text{in}}$ and $\Nt \Nrf \mathcal{I}_{\text{AO}}$ respectively, while that of the proposed ManNet-based design is only $\mathcal{I}_{\text{net}} L$. In general, both $\mathcal{I}_{\text{net}}$ and $L$ are of the same order as $\Nrf$, and thus, $\mathcal{I}_{\text{net}} L \ll \Nt \Nrf \mathcal{I}_{\text{AO}}$ and $\mathcal{I}_{\text{net}} L \ll \mathcal{I}_{\text{MO}}^{\text{in}} \mathcal{I}_{\text{MO}}^{\text{out}}$. For example, in a simulation with $\Nt=128$, $\Nr=\Nrf=\Ns=2$, and $K=128$, we found that $\mathcal{I}_{\text{net}}=10$ and $L=7$ are sufficient for ManNet-based FC-HBF to achieve satisfactory performance, whereas AO and MO-AltMin require up to $\Nt \Nrf \mathcal{I}_{\text{AO}} = 250$ and $\mathcal{I}_{\text{MO}}^{\text{out}} \mathcal{I}_{\text{MO}}^{\text{in}} = 500$ iterations to converge, respectively (this will be shown in Section \ref{sec_sim}, Fig.\ \ref{fig_rate_conv}). Therefore, the proposed algorithm performs much faster than MO-AltMin, and its computational complexity is considerably lower than MO-AltMin and AO, and comparable with that of OMP.
	
	
	\section{Proposed SC-HBF Designs}
	\label{sec_SCHBF}
	Next, we present the deep unfolding based dynamic SC-HBF design. As the fixed SC-HBF architecture is  a special case of the dynamic one, below we present the general solution to the latter. We first consider the following problem:
	\begin{subequations}
		\label{opt_prob_RF_sub}
		\begin{align} 
			\quad \underset{\substack{ \Frf, \{\Fbb[k]\} }}{\textrm{minimize}}\ & \sum_{k=1}^{K}\norm{\bF_{\text{opt}}[k] - \Frf \Fbb[k]}_{\mathcal{F}}^2,\\
			\textrm{subject to}\ &\Frf \in \Asub. \nbthis \label{cons_sub}
		\end{align}
	\end{subequations}
	Compared to the FC-HBF design in \eqref{opt_prob_RF}, problem \eqref{opt_prob_RF_sub} inherits the nonconvexity due to the unit-modulus constraint of the nonzero analog precoding coefficients. Furthermore, unlike the cases of FC-HBF and fixed SC-HBF, the connections between the RF chains and antennas are also design variables in this problem. The joint optimization of the RF chain-antenna connections, $\Frf$, and $\Fbb[k]$ is challenging. Herein we propose efficient algorithms to solve \eqref{opt_prob_RF_sub} with the main idea being to decouple the design variables.
	
	\subsection{ManNet-based Heuristic FC-HBF Design}
	\label{sec_connections}
	Let $\mC \in \setN^{\Nt \times \Nrf}$ denote the mapping matrix defining the connections between the $\Nrf$ RF chains and $\Nt$ antennas such that
	\begin{align*}
		&[\mC]_{m,n} = 
		\begin{cases*}
			1, \text{if $[\Frf]_{m,n} \neq 0$}\\
			0, \text{otherwise}
		\end{cases*},\ \forall m, n, \nbthis \label{cons_bin} \\
		&\sum_{m=1}^{\Nt} [\mC]_{m,n} = M,\ \forall n, \nbthis \label{cons_col} \\
		&\sum_{n=1}^{\Nrf} [\mC]_{m,n} = 1,\ \forall m. \nbthis \label{cons_row}
	\end{align*}
	With the introduction of variable $\mC$, the dynamic SC-HBF optimization can be rewritten as
	\begin{subequations}
		\label{opt_prob_RF_sub_C}
		\begin{align*} 
			\quad \underset{\substack{ \mC, \Frft, \{\Fbb[k]\} }}{\textrm{minimize}}\ & \sum_{k=1}^{K}\norm{\bF_{\text{opt}}[k] - (\mC \odot \Frft) \Fbb[k]}_{\mathcal{F}}^2, \nbthis \label{obj_sub_C} \\
			\textrm{subject to}\ &\Frft \in \Afull, \nbthis \label{cons_sub_Ftilde}\\
			&\eqref{cons_bin}-\eqref{cons_row}. \nbthis \label{cons_sub_C}
		\end{align*}
	\end{subequations}
	Note that in this problem, the sub-connected structure constraint on the analog precoder, i.e., \eqref{cons_sub}, has been relaxed, as seen in \eqref{cons_sub_Ftilde}. This efficiently decouples the designs of the RF chain/antenna connections and the analog precoder. Because $\mC$ is a matrix of binary entries, its optimal solution could be found by exhaustive search over all possibilities, but with a prohibitive complexity (exponential in $\Nt \Nrf$). To avoid this, we investigate the achievable SE of the analog precoders given as $R_{\text{RF}} = \frac{1}{K} \sum_{k=1}^{K} R_{\text{RF},k}$, where
	\begin{align*}
		R_{\text{RF},k} &= \log_2 \text{det} \Big( \bI_{\Nr} + \frac{\rho}{\sigma^2_{\text{n}} \Ns} \mH[k] (\mC \odot \Frf) \\
		&\hspace{3cm} \times (\mC \odot \Frf)^H \mH[k]^H \Big).  \nbthis \label{eq_SE_analog}
	\end{align*}
	It is observed that for a given $\mH[k]$, to achieve the highest signal-to-noise ratio (SNR), $\mC$ should be designed to match the nonzero entries in $\Frf$ with the ``best'' coefficients of $\mH[k]$, i.e., those with the largest absolute values. Based on this observation, we propose Algorithm \ref{alg_mapping} to determine $\mC$ for any $\mH[k]$. Furthermore, because of the relaxation in \eqref{cons_sub_Ftilde}, ManNet can be used to produce $\Frft \in \Afull$. Then, for each $\mH[\tilde{k}]$, with $\tilde{k} \in \tilde{\mathcal{K}} \subseteq \{1,2,\ldots,K\}$, $\mC$ is determined using Algorithm \ref{alg_mapping}, and the $\Fbb[k]$ are found using \eqref{eq_Fbb}. The final solutions for $\Frf$ and $\{\Fbb[k]\}$ are those that provide the best performance, i.e., the largest SE. This heuristic ManNet-based SC-HBF approach is summarized in Algorithm \ref{alg_ManNet_SC_HBF}. 
	
	We note that although the proposed ManNet-based SC-HBF scheme can avoid an exhaustive search for $\mC$ for each channel $\mH[k]$, it still requires $\absshort{\tilde{\mathcal{K}}}$ iterations to obtain $\Frf^{(\tilde{k})}$ and  $\{\Fbb[k]\}^{(\tilde{k})},\ (\tilde{k} \in \tilde{\mathcal{K}})$. We will show later that such an iterative process yields very satisfactory performance for SC-HBF, at the expense of increased complexity and run time.
	
	\begin{algorithm}[t]
		\small
		\caption{Dynamic RF chain - antenna Mapping}
		\label{alg_mapping}
		\begin{algorithmic}[1]
			\REQUIRE $\mH[k]$.
			\ENSURE $\mC$ satisfying \eqref{cons_bin}-\eqref{cons_row}.
			\STATE Set $\tilde{\mH}[k]$ to the matrix containing $\Nrf$ rows of $\mH[k]$ with largest norm values. Obtain $\bar{\mH}$ such that $[\bar{\mH}]_{i,j} = \abs{[\tilde{\mH}[k]]_{i,j}}, \forall i, j$.
			\STATE Set $\mC$ with $[\mC]_{m,n} = 1, \forall m, n$.
			\FOR{$m = 1 \rightarrow M$}
			\FOR{$n = 1 \rightarrow \Nrf$}
			\STATE Set $m_0$ to the index of the smallest element in the $n$-th column of $\bar{\mH}$.
			\STATE Set $[\mC]_{m_0,n} = 0$. 
			\STATE Set all elements in the $m_0$-th row of $\bar{\mH}$ to zeros.
			\ENDFOR
			\ENDFOR
		\end{algorithmic}
	\end{algorithm} 
	
	\begin{algorithm}[t]
		\small
		\caption{Heuristic ManNet-based SC-HBF}
		\label{alg_ManNet_SC_HBF}
		\begin{algorithmic}[1]
			\REQUIRE $\mH, \Fopt$, and the trained ManNet.
			\ENSURE $\Frf, \{\Fbb[k]\}$.
			\STATE Apply Algorithm \ref{alg_ManNet_HBF} to obtain $\Frft  \in \Afull$.
			\FOR{$\tilde{k} \in \tilde{\mathcal{K}}$}
			\STATE Obtain $\mC^{(\tilde{k})}$ for $\mH[\tilde{k}]$ using Algorithm \ref{alg_mapping}.
			\STATE Obtain $\Frf^{(\tilde{k})} = \mC^{(\tilde{k})} \odot \Frft$.
			\STATE Solve $\{\Fbb[k]\}^{(\tilde{k})}$ using \eqref{eq_Fbb}.
			\ENDFOR
			\STATE Return $\Frf$ and $\{\Fbb[k]\}$ that provide the largest SE.
		\end{algorithmic}
	\end{algorithm}

	\subsection{Low-Complexity subManNet-based SC-HBF}
	\label{sec_subManNet}
	Here we propose a computationally efficient SC-HBF design to avoid the iterative procedure as well as the extra complexity to produce $\Frft \in \Afull$, as done in Algorithm \ref{alg_ManNet_SC_HBF}. This can be achieved if a good channel is chosen in advance to design $\mC$, and if the employed DNN only generates the nonzero coefficients of $\Frf \in \Asub$. These assumptions motivate a subcarrier selection scheme and the design of subManNet, a simplified version of ManNet proposed below.
	
	\subsubsection{Subcarrier Selection}
	First, we observe from \eqref{eq_SE_analog} that the transmissions via different subcarriers have different contributions to the total achievable SE. Specifically, let $R_{\text{RF},k^{\star}}$ be the maximum SE of all the sub-carriers, i.e., $R_{\text{RF},k^{\star}} = \max \{R_{\text{RF},1}, \ldots, R_{\text{RF},K}\}$. Then, $R_{\text{RF},k^{\star}}$ has the most significant contribution to $R_{\text{RF}}$. On the other hand, for any given $\Frf \in \Asub$, the $\Fbb[k]$ can be optimally found using the closed-form solution in \eqref{eq_Fbb}. These observations motivate us to design $\mC$ to maximize $R_{\text{RF},k^{\star}} = \log_2 \text{det} ( \bI_{\Nr} + \frac{\rho}{\sigma^2_{\text{n}} \Ns} \mH[k^{\star}] (\mC \odot \Frf) (\mC \odot \Frf)^H \mH[k^{\star}]^H )$. Here, because of the unity-modulus constraints on the non-zero elements of $\Frf$, subcarrier $k^{\star}$ is chosen such that the channel $\mH[k^{\star}]$ has the largest Frobenius norm among all the channels. Thus, $\mC$ is determined based on $\mH[k^{\star}]$ using Algorithm \ref{alg_mapping}.
	
	\subsubsection{subManNet-based SC-HBF}
	\label{sec_subManNet_SCHBF}
	Once $\mC$ is determined, let
	\begin{align*}
		\vc = \mathcal{V}(\mC + j \mC) \in \setR^{N \times 1}, \nbthis \label{eq_c}
	\end{align*}
	where $\mathcal{V}$ is defined in \eqref{def_transformation}. By similar transformations as in \eqref{eq_trans_x}--\eqref{eq_trans_obj_real}, we can rewrite the objective  of problem \eqref{opt_prob_RF_sub} as
	\begin{align*}&\sum_{k=1}^{K}\norm{\bF_{\text{opt}}[k] - \Frf \Fbb[k]}_{\mathcal{F}}^2\\
		&\hspace{1cm}= \sum_{k=1}^{K}\norm{\bF_{\text{opt}}[k] - (\mC \odot \Frft) \Fbb[k]}_{\mathcal{F}}^2 \\
		&\hspace{1cm}= \sum_{k=1}^{K} \norm{{\vz[k]} - {\bB[k]} (\vc \odot \vx)}^2. \nbthis \label{eq_trans_obj_real}
	\end{align*}
	Problem \eqref{opt_prob_RF_sub} is then transformed to
	\begin{align*}
		\vx^{\star} = \argmin_{\vx: \mathcal{V}^{-1} (\vx) \in \Asub} \sum_{k=1}^{K} \norm{\vz[k] - \bB[k] (\vc \odot \vx)}^2. \nbthis \label{eq_opt_sol_sub}
	\end{align*}
	This motivates us to specialize ManNet for SC-HBF design.
	
	Specifically, we propose subManNet to learn and output $\vx^{\star}$ in \eqref{eq_opt_sol_sub}. In subManNet, the activation function is set to
	\begin{align*}
		\sigma_{t,\vc}(\vx) = \vc \odot \psi_t(\vx), \nbthis \label{eq_activation_function_sub}
	\end{align*}
	where $\psi_t(\cdot)$ is defined in \eqref{def_activation} and $\vu_{\ell-1}$ is modified as
	\begin{align*}
		\tilde{\vu}_{\ell-1} = \vc \odot \vu_{\ell-1}. \nbthis \label{eq_u_mask}
	\end{align*}
	As a result, the $n$-th nodes in both the sub-networks associated with input vectors $\vx_{\ell-1}$ and $\Tilde{\vu}_{\ell-1}$ do not require any computations if $c_n = 0$. In other words, subManNet produces the output based on the predetermined RF chain/antenna connections specified in $\mC$. The offline training and online application of subManNet can be performed similarly to ManNet, except for the aforementioned modifications. We omit the detailed training process here but summarize the proposed subManNet-based SC-HBF design in Algorithm \ref{alg_subManNet_HBF}. Its first step is to design the mapping matrix $\mC$ for the best channel $\mH[k^{\star}]$, and the remaining process is similar to Algorithm \ref{alg_ManNet_HBF}, except for the pre-processing of $\vu_{\ell - 1}$. We outline the structure of subManNet in Fig.\ \ref{fig_iterative_unfolding}(c).
	
	\subsection{Complexity Analysis}
	\label{sec_complexity_sub}
	In Algorithm \ref{alg_ManNet_SC_HBF}, each iteration is performed with a complexity of $\mathcal{O}( 2 \Nt\Nr \Nrf)$. This is mainly to solve $\{\Fbb[k]\}^{(\tilde{k})}$ with \eqref{eq_Fbb}, while steps 3 and 4 require very few computations. Thus, we approximate the total complexity of Algorithm \ref{alg_ManNet_SC_HBF} as
	\begin{align*}
		&\mathcal{C}_{\textrm{ManNet-SCHBF}}^{\text{heuristic}} = \mathcal{C}_{\textrm{ManNet-FCHBF}} +  \absshort{\tilde{\mathcal{K}}} \mathcal{O}( 2\Nt\Nr \Nrf). \nbthis \label{eq_comp_ManNetHBF_ES}
	\end{align*}
	
	On the other hand, subManNet offers a complexity reduction by a factor of $\Nrf$ compared to ManNet. This is consistent with the requirement of $\Nrf$ times fewer phase shifters in the sub-connected architecture. Thus, the overall complexity of the subManNet-based SC-HBF scheme in Algorithm \ref{alg_subManNet_HBF} is
	\begin{align*}
		&\mathcal{C}_{\textrm{subManNet-SCHBF}}
		= \left(\mathcal{I}_{\text{net}} - 1\right) \mathcal{O} (\Nt K  \Nrf^2 ) + \mathcal{O} (\Nt  K \Nrf )  \\ 
		&\qquad + \mathcal{I}_{\text{net}} \mathcal{O} (2 K \Nrf^2 \Ns + L (3\Nt + 2K \Ns) ), \nbthis \label{eq_comp_sub_ManNetHBF}
	\end{align*}
	based on the complexity analysis of the ManNet-based FC-HBF scheme in Section \ref{sec_comp_analysis_ManNet_HBF}. In particular, subManNet inherits the fast convergence and low complexity of ManNet, i.e., it only requires small $\mathcal{I}_{\text{net}}$ and $L$ to achieve good performance. SDR-AltMin \cite{yu2016alternating} requires complexities of $\mathcal{O}(K \Nt \Ns)$ and $\mathcal{O}(K\Ns^3 \Nrf^3)$ to obtain the analog and digital precoders, respectively, in each iteration. Thus, its total complexity is $\mathcal{C}_{\textrm{SDR-AltMin}} = \mathcal{I}_{\text{SDR}} \mathcal{O}(\Nt K \Ns +  K \Ns^3 \Nrf^3)$, where $\mathcal{I}_{\text{SDR}}$ is the number of iterations for alternating updates of $\Frf$ and $\Fbb[k]$. Our simulations will show that the proposed design also performs better and much faster than SDR-AltMin. 
	
	Based on the fact that $\Nt, K \gg \Nrf, \Ns, \Nr, L$, we present the approximate complexities of the discussed approaches in Table \ref{tab_complexity} to facilitate complexity comparisons. It can be seen that the proposed deep unfolding schemes, OMP, and SDR-AltMin have comparable complexities, which are all much lower than those of AO and MO-AltMin. In particular, the complexity of AO increases exponentially with $\Nt$.
	
	\renewcommand{\arraystretch}{1.3}
	\begin{table}[t!]
		\small
		\begin{center}
			\caption{Computational complexity of ManNet/subManNet based FC-HBF/SC-HBF  compared with  MO-AltMin, AO, OMP, and SDR-AltMin.}
			\label{tab_complexity}
			\begin{tabular}{|c|c|c|c|}
				\hline
				Structure & Schemes & Overall complexity\\
				\hline
				\hline
				\multirow{4}{*}{FC-HBF}
				&{ManNet} & {$\mathcal{I}_{\text{net}} \mathcal{O}(\Nt K + \Nt + K)$}\\
				\cline{2-3}
				&MO-AltMin  & $\mathcal{I}_{\text{MO}}^{\text{out}} \mathcal{O}(\Nt K) + \mathcal{I}_{\text{MO}}^{\text{out}} \mathcal{I}_{\text{MO}}^{\text{in}} \mathcal{O}(\Nt + K)$  \\
				\cline{2-3}
				&AO  & $\mathcal{O}(\Nt K) + \mathcal{I}_{\text{AO}} \mathcal{O}(\Nt^2)$ \\
				\cline{2-3}
				&OMP  & $\mathcal{O}(\Nt K + \Nt)$\\
				\hline
				\hline
				\multirow{3}{*}{SC-HBF}
				&{ManNet} & {$\mathcal{I}_{\text{net}} \mathcal{O}(\Nt K + \Nt + K) + 2 \absshort{\tilde{\mathcal{K}}} \mathcal{O}( \Nt)$}\\
				\cline{2-3}
				&{subManNet} & {$\mathcal{I}_{\text{net}} \mathcal{O}(\Nt K + \Nt + K)$}\\
				\cline{2-3}
				&SDR-AltMin  & $\mathcal{I}_{\text{SDR}} \mathcal{O}(\Nt K +  K)$\\
				\hline
			\end{tabular}
		\end{center}
	\end{table}
	
	\begin{algorithm}[t]
		\small
		\caption{subManNet-based SC-HBF}
		\label{alg_subManNet_HBF}
		\begin{algorithmic}[1]
			\REQUIRE $\mH, \Fopt$, ManNet's trained parameters $\left\{ \{\vw_{\ell,1}, \vw_{\ell,2}\}_{\ell=1}^L, t \right \}$.
			\ENSURE $\Frf, \{\Fbb[k]\}$.
			\STATE Apply Algorithm \ref{alg_mapping} for $\mH[k^{\star}] = \max\{\mH[1], \ldots, \mH[K]\}$ to obtain RF chain - antenna mapping matrix $\mC$.
			\STATE Apply Algorithm \ref{alg_ManNet_HBF} with $\psi_t(\hat{\bx}_{\ell}^{(i)})$ and $\vu_{\ell-1}$ replaced by $\sigma_{t,\vc}(\vx)$ and $\tilde{\vu}_{\ell-1}$ in \eqref{eq_activation_function_sub} and \eqref{eq_u_mask}, respectively, to obtain $\Frf \in \Asub$ and $\{\Fbb[k]\}$.
		\end{algorithmic}
	\end{algorithm}

	\section{Simulation Results}
	\label{sec_sim}
	
	\begin{figure}[t]
		\includegraphics[scale=0.58]{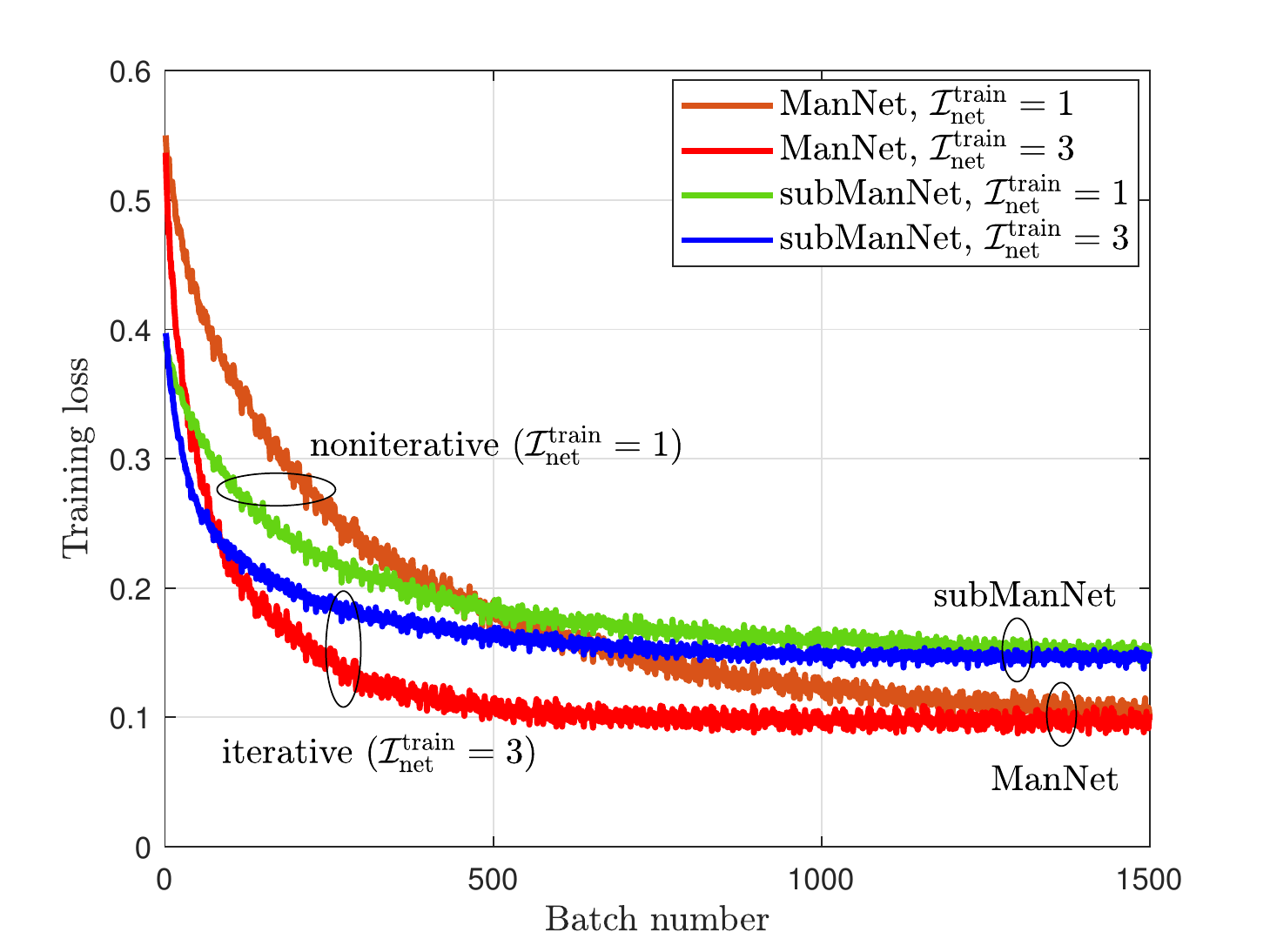}
		\caption{Normalized training loss of ManNet and subManNet with $\Nt = 64$, $K=128$, $\Nr=\Nrf=\Ns = 2$, $L = 6$, and $\mathcal{I}_{\text{net}}^{\text{train}}=\{1, 3\}$.}
		\label{fig_loss}
	\end{figure}
	In this section, we provide numerical results to demonstrate the performance of the proposed deep unfolding solutions for FC-HBF and SC-HBF designs. We first detail the simulation setup and  benchmarks, after which we discuss the results in terms of SE and complexity. 
	
	\subsection{Simulation Setup and Training of DNNs}
	We assume scenarios with $\Nt = \{16,32,64,128\}$, $K=128$, and $\Nr = \Nrf = \Ns = 2$. The channel realizations are generated based on \eqref{eq_channel_model} with $P = 4$, $\phi^\text{t}_{p}, \phi^\text{r}_{p} \sim \mathcal{U} [0^\circ,360^\circ)$, $\theta^\text{t}_{p}, \theta^\text{r}_{p} \sim \mathcal{U} [-90^\circ, 90^\circ]$, $\alpha_{p} \sim \mathcal{CN}(0, 1)$ \cite{yu2016alternating}, {and $\tau_p \sim \mathcal{U}[0,\tau_{\max}]$, where $\tau_{\max} = Q T_{\text{s}}$ with $T_{\text{s}}$ being the sampling period and $Q$ being the cyclic prefix length, which is set to $\frac{K}{4}$ similar to IEEE 802.11ad \cite{alkhateeb2016frequency, park2017dynamic}}. The center frequency and bandwidth are set to $f_{\text{c}} = 300$ GHz and BW $= 30$ GHz, respectively. ManNet and subManNet are implemented using Python with the Pytorch library and a Tesla V100-SXM2 processor. {For the training phase, a learning rate of $0.0001$ is used with the Adam optimizer, and we set $L = \{4,5,6,7\}$ and $|\mathcal{D}|=\{400, 500, 600, 700\}$ for $\Nt = \{16,32,64,128\}$, respectively.} The SNR is defined as SNR $ = \rho/\sigma_{\text{n}}^2$. The  results are averaged over $100$ iterations.
	
	We first show the loss obtained in \eqref{eq_loss} during training ManNet and subManNet with $\Nt = 64$ and $\Nr = \Nrf = \Ns = 2$ in Fig.\ \ref{fig_loss}. Both networks are trained using Algorithm \ref{alg_train_ManNet}, but the latter employs the modified activation function \eqref{eq_activation_function_sub} and input vector \eqref{eq_u_mask}, as discussed earlier in Section \ref{sec_subManNet}. We consider  $\mathcal{I}_{\text{net}}^{\text{train}}=\{1, 3\}$, corresponding to the non-iterative and iterative training approaches, respectively. It is seen for both the DNNs that the loss decreases and essentially converges, but at different speeds and to different values. Specifically, it is clear that with $\mathcal{I}_{\text{net}}^{\text{train}} = 3$, the DNNs converge rapidly after about $800$ batches. In contrast, when the non-iterative training is applied, they converge more slowly, and convergence has not been reached even after $1500$ batches. Because the objective $\sum_{k=1}^{K}\norm{\bF_{\text{opt}}[k] - \Frf \Fbb[k]}^2_{\mathcal{F}}$ attained by FC-HBF is smaller than that of SC-HBF, it is reasonable that the converged loss of ManNet is smaller than that of subManNet. As the loss function \eqref{eq_loss} also measures the objective in \eqref{opt_prob_RF} and \eqref{opt_prob_RF_sub}, the convergence of the training loss reflects the abilities of ManNet and subManNet to solve problems \eqref{opt_prob_RF} and \eqref{opt_prob_RF_sub}, respectively. Note that in Fig.\ \ref{fig_loss}, the training loss is with respect to the total number of batches over all training epochs. Equivalently, the training losses for the iterative and non-iterative schemes have converged within $30$ and $50$ epochs, respectively. 
	
	\subsection{Performance of Proposed Deep Unfolding HBF Schemes}
	\label{sec_sim_performance}
	
	\begin{figure}[t]%
		\centering
		\subfigure[SNR $= 10$ dB]{%
			\label{fig_rate_conv_10dB}
			\includegraphics[scale=0.6]{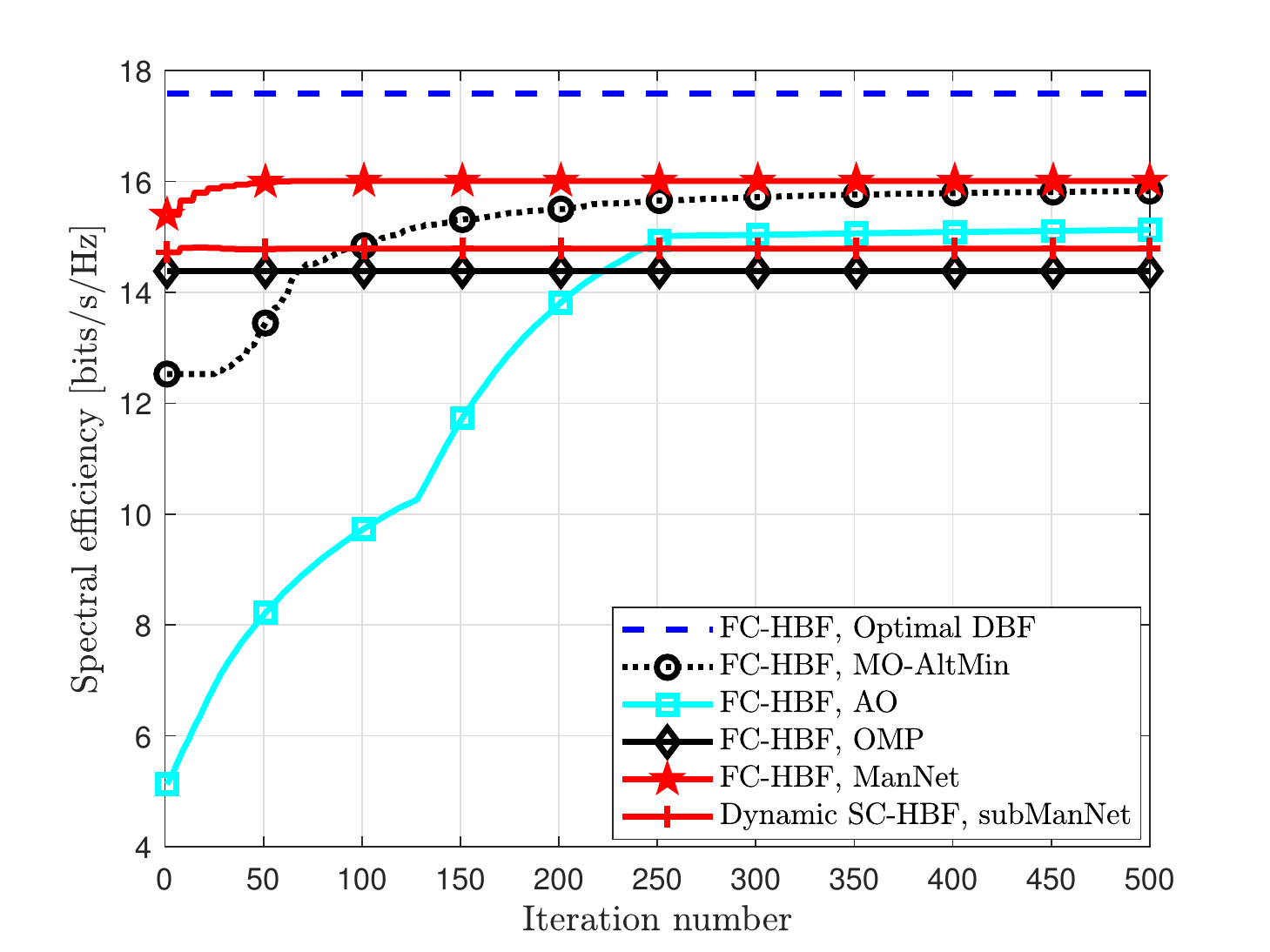}}\\
		\subfigure[SNR $= 20$ dB]{%
			\label{fig_rate_conv_20dB}
			\includegraphics[scale=0.6]{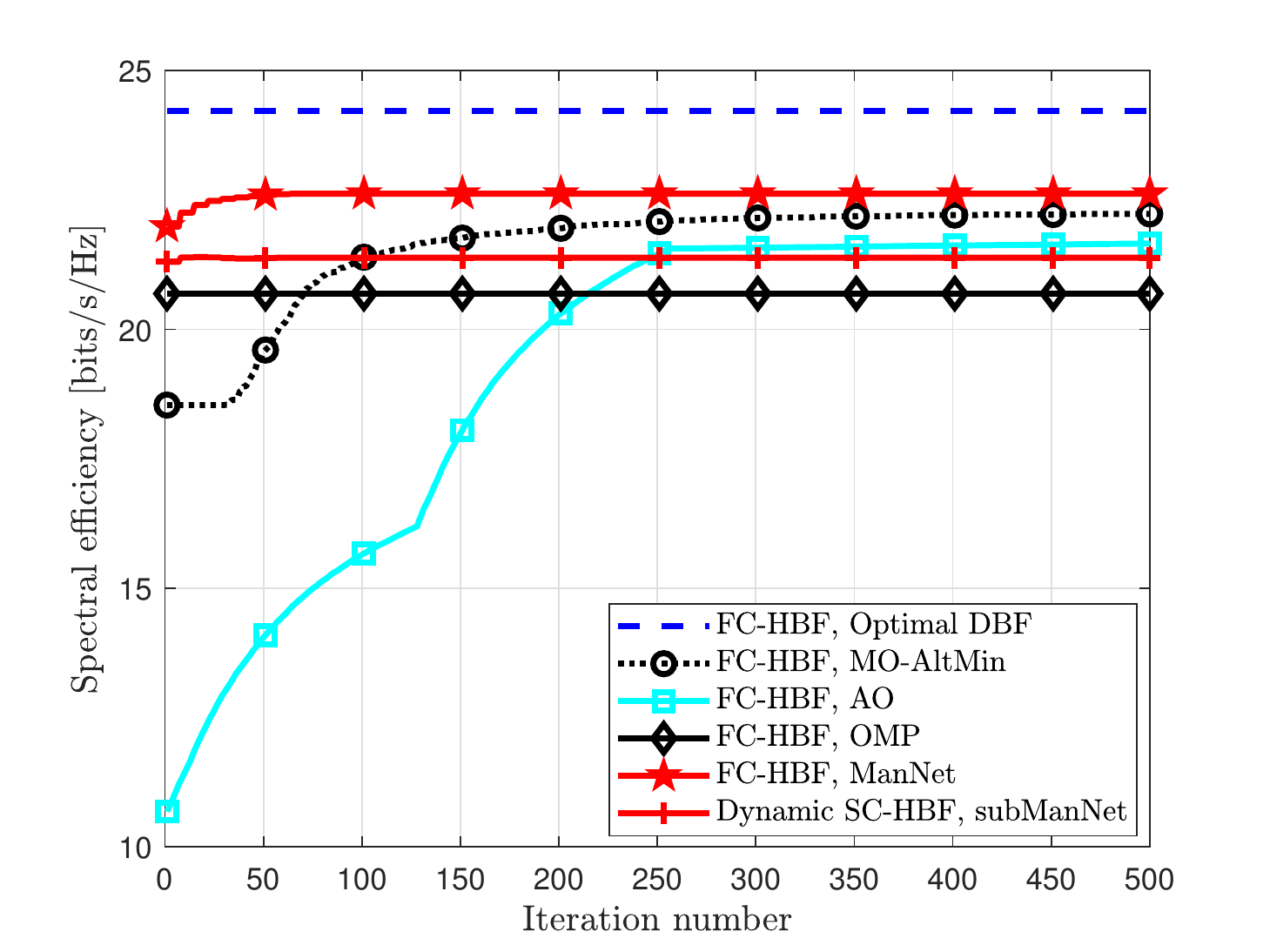}} 
		\caption[]{Convergence of ManNet and subManNet-based HBF with $\Nt = 128$, $\Nr=\Nrf=\Ns = 2$, and SNR $=\{10, 20\}$ dB.}%
		\label{fig_rate_conv}%
	\end{figure}
	
	Here, we investigate the performance of the proposed deep unfolding FC-HBF and SC-HBF designs based on ManNet and subManNet in their online applications, i.e., in Algorithms \ref{alg_ManNet_HBF}--\ref{alg_subManNet_HBF}. We train the DNNs over $\mathcal{I}_{\text{net}}^{\text{train}} = 3$ iterations. For comparisons of FC-HBF designs with ManNet in Algorithm \ref{alg_ManNet_HBF}, we consider optimal fully digital beamforming (DBF), MO-AltMin \cite{yu2016alternating}, OMP \cite{el2014spatially, lee2014af}, and AO \cite{nguyen2022beam}. The dynamic SC-HBF designs with ManNet in Algorithm \ref{alg_ManNet_SC_HBF} and with subManNet in Algorithm \ref{alg_subManNet_HBF} are compared with the SDR-AltMin scheme \cite{yu2016alternating}.
	
	In Fig.\ \ref{fig_rate_conv}, we compare the convergence of the considered schemes with $\Nt=128$, $\Nr = \Nrf = \Ns = 2$, $K=128$, SNR $=\{10,20\}$ dB, $L=7$, and $\mathcal{I}_{\text{net}} = 10$. We note that OMP and optimal DBF are not iterative, so their performance is constant over the number of iterations. Among the iterative schemes, MO-AltMin converges the slowest, and it has not strictly converged after $500$ iterations. AO converges faster than MO-AltMin, but still requires about $250$ iterations and converges to unsatisfactory performance. In contrast, the performance of the proposed ManNet-based FC-HBF and subManNet-based SC-HBF methods improves rapidly and reaches satisfactory values after only tens of iterations. Particularly, among the sub-optimal schemes, ManNet-HBF achieves the highest SE. It is observed that the SE of ManNet-HBF increases step-by-step, over $\mathcal{I}_{\text{net}} = 10$ steps, and reaches its maximum after $\mathcal{I}_{\text{net}} L = 70$ iterations. This is because $L=7$ layers is the number of inner iterations used to perform steps 6--10 in Algorithm \ref{alg_ManNet_HBF}, and in these layers the performance does not improve. However, because the weights of the DNNs are applied once to generate the output, the maximum SE of ManNet and subManNet is reached after only $\mathcal{I}_{\text{net}}$ iterations. This figure clearly shows the advantages of the proposed scheme in accelerating HBF transceiver design and optimization.
	
	\begin{figure}[t]%
		\centering
		\subfigure[Fully-connected HBF architecture]{%
			\label{fig_SE_SNR_FC}
			\includegraphics[scale=0.6]{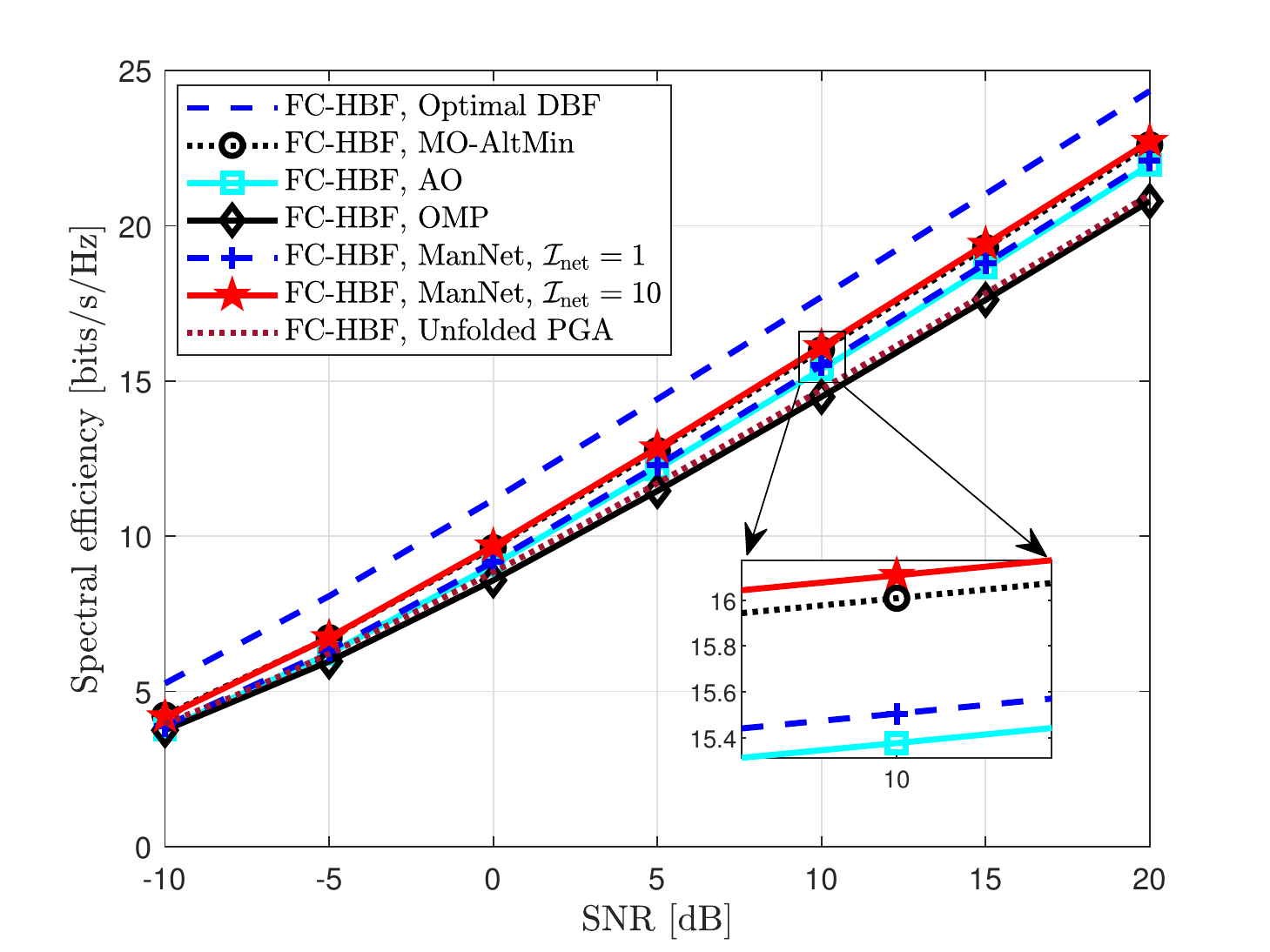}}\\
		\vspace{-0.3cm}\subfigure[Sub-connected HBF architecture]{%
			\label{fig_SE_SNR_SC}
			\includegraphics[scale=0.6]{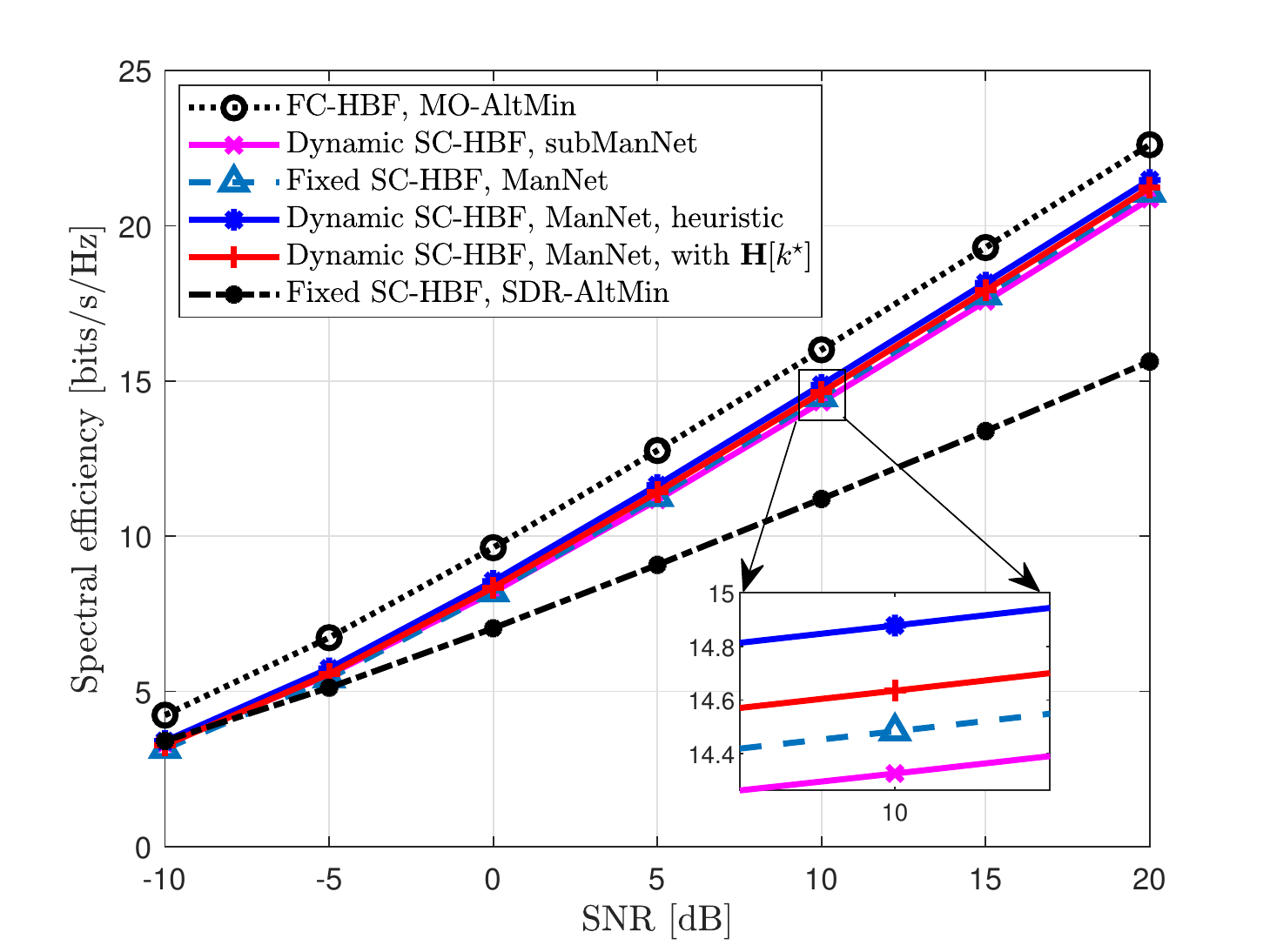}} 
		\caption[]{SE performance of the proposed ManNet-HBF designs for FC-HBF and SC-HBF with $\Nt=128$, $\Nr = \Nrf = \Ns = 2$, and $K=128$.}%
		\label{fig_SE_SNR}%
	\end{figure}
	
\begin{figure}[t]
		\includegraphics[scale=0.58]{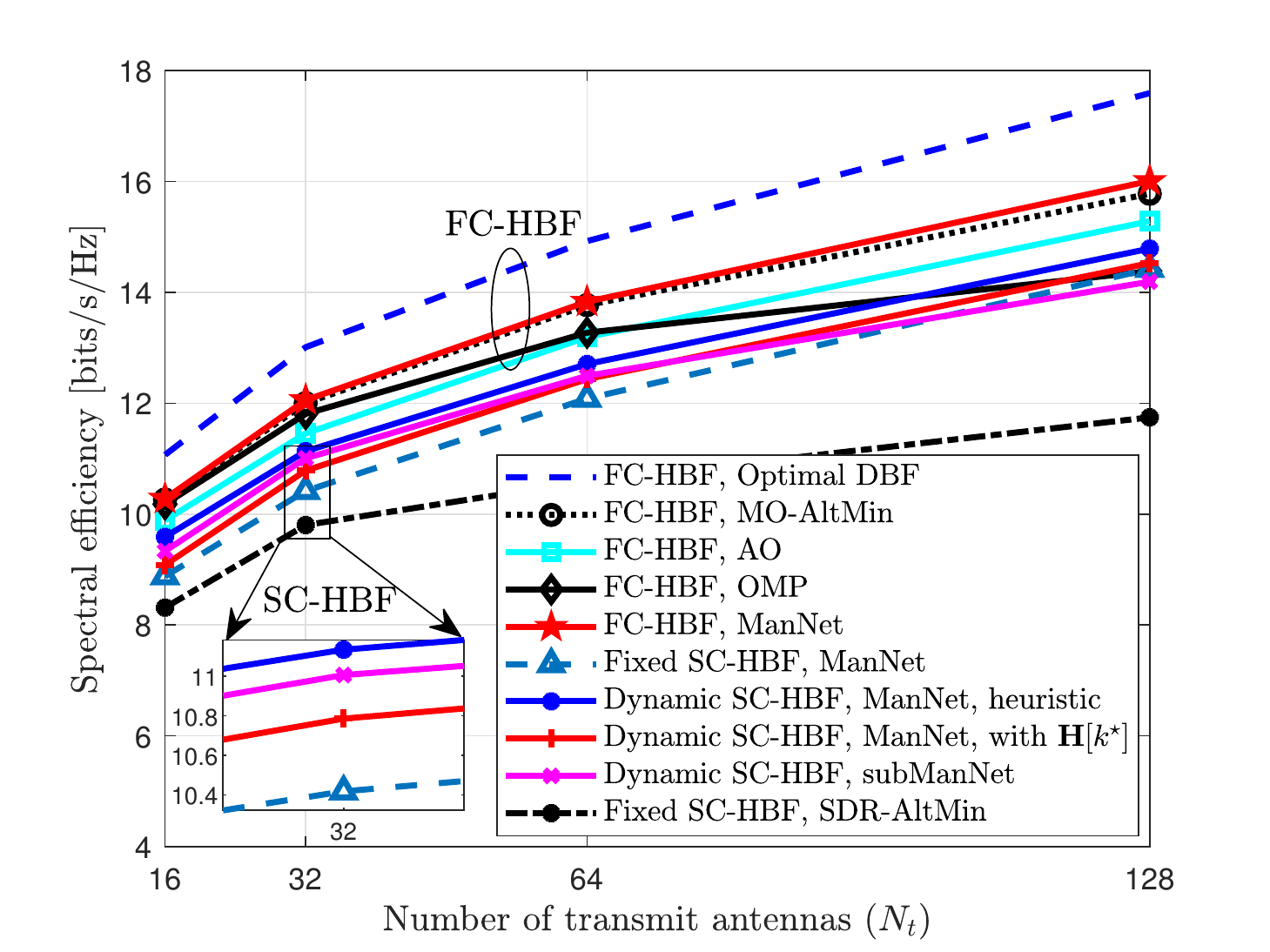}
		\caption{SE performance of ManNet and subManNet-based HBF schemes with $\Nt \in [16,128]$, $\Nr=\Nrf=\Ns = 2$, and SNR $=10$ dB.}
		\label{fig_SE_Nt}
	\end{figure}

	In Figs.\ \ref{fig_SE_SNR} and \ref{fig_SE_Nt}, we compare the SE performance attained by the proposed deep unfolding schemes, including ManNet- and subManNet-based FC-HBF and SC-HBF in Algorithms \ref{alg_ManNet_HBF}--\ref{alg_subManNet_HBF}, with that of the optimal DBF, MO-AltMin, AO, OMP, SDR-AltMin, and the unfolded PGA approach \cite{agiv2022learn} with $5$ iterations. Furthermore, we also show the performance of the dynamic SC-HBF based on ManNet without a heuristic search for $\mC$ (referred to as \textit{``Dynamic SC-HBF, ManNet, with $\mH[k^{\star}]"$} in the figures). In addition, we present the results for the fixed SC-HBF scheme based on ManNet (referred to as \textit{``Fixed SC-HBF, ManNet"}), i.e., in which $\mC$ is fixed to $\mC = \mathrm{blkdiag}\{\mathbf{1}_{M}, \ldots, \mathbf{1}_{M}\}$, where $\mathbf{1}_{M}$ denotes a column vector of $M$ ones. 
	
	In Fig.\ \ref{fig_SE_SNR}, we set $\Nt=128, \Nr = \Nrf = \Ns = 2$, and $K=128$. The convergence tolerance is set to $10^{-3}$ for the iterative MO-AltMin, AO, and SDR-AltMin approaches, and $\mathcal{I}_{\text{net}} = \{1, 10\}$ is set for the ManNet-based FC-HBF scheme. Note that for $\mathcal{I}_{\text{net}} = 1$, $\Frf$ is obtained directly using ManNet without an iterative update, and the $\Fbb[k]$ are solved directly using \eqref{eq_Fbb}. For the heuristic ManNet-based SC-HBF scheme in Algorithm \ref{alg_ManNet_SC_HBF}, we use $\tilde{\mathcal{K}} = \{1, 3, 5, \ldots, K-1\}$. From Fig.\ \ref{fig_SE_SNR}, the following observations are made:
	\begin{itemize}
		\item In Fig.\ \ref{fig_SE_SNR_FC}, FC-HBF based on ManNet with $\mathcal{I}_{\text{net}} = 10$ performs better than MO-AltMin and AO, and much better than five unfolded PGA iterations and OMP, even with only $\mathcal{I}_{\text{net}} = 1$ iteration. At SNR $=10$ dB, the proposed ManNet-based FC-HBF scheme with $\mathcal{I}_{\text{net}} = 10$ achieves $90.95\%$ of the optimal performance, while the performance of MO-AltMin, AO, unfolded PGA, and OMP are only at $90.38\%$, $86.82\%$, $83.16\%$ and $81.82\%$ of the optimum, respectively.
		
		\item The heuristic dynamic SC-HBF design based on ManNet (i.e., Algorithm \ref{alg_ManNet_SC_HBF}) provides superior performance, as seen in Fig.\ \ref{fig_SE_SNR_SC}. The other deep unfolding SC-HBF schemes perform slightly worse than the heuristic one, but they all outperform SDR-AltMin for SNR $\geq -5$ dB. At SNR $=10$ dB, the proposed deep unfolding SC-HBF schemes achieve $90-93\%$ of the FC-HBF performance based on MO-AltMin, while that achieved by SDR-AltMin is only at $70\%$.
		
		\item SC-HBF designs based on ManNet perform better than that with subManNet. This is reasonable since the fully-connected analog precoder produced by ManNet is more reliable than the sub-connected version, as observed from Fig.\ \ref{fig_loss}. The dynamic ManNet-based SC-HBF algorithm performs just slightly better than the fixed version. We note here that larger gains can be attained with smaller $\Nt$, as will be shown next.
	\end{itemize} 
	
	In Fig.\ \ref{fig_SE_Nt}, we plot the SE performance of the considered schemes for $\Nt=\{16,32,64,128\}$, $\Nr = \Nrf = \Ns = 2$, $K=128$, SNR = $10$ dB, and $\mathcal{I}_{\text{net}} = 10$. It is observed that OMP only performs well for small $\Nt$ and has significant performance loss as $\Nt$ increases. Among the sub-optimal FC-HBF schemes, the proposed ManNet-HBF approach achieves the best performance, which is slightly better than MO-AltMin and far better than AO and OMP for all considered $\Nt$. Comparing the SC-HBF schemes, the heuristic ManNet-based SC-HBF design has the best performance. The subManNet-based SC-HBF algorithm performs very close to the heuristic one for $\Nt \leq 64$. Furthermore, it is seen that compared to fixed SC-HBF, the gains achieved from dynamic SC-HBF are more significant for small and moderate $\Nt$. This is reasonable because as $\Nt$ increases, all the sub-arrays become large and the beamforming gain is guaranteed even without the optimized connections between RF chains and antennas.
	
	\subsection{Computational and Time Complexity Comparison}
	
\begin{figure}[t]
		\includegraphics[scale=0.58]{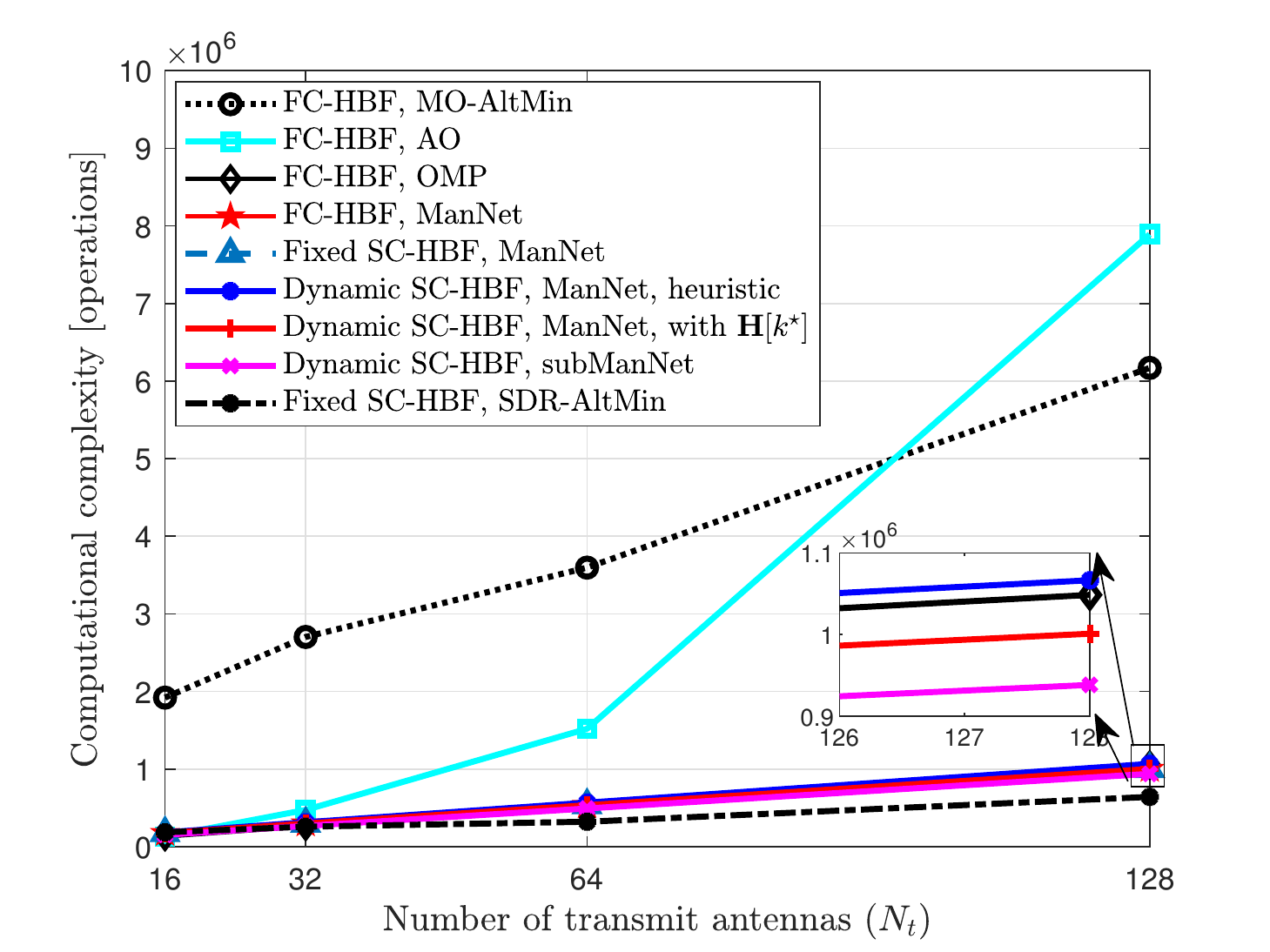}
		\caption{Computational complexity of ManNet and subManNet-based HBF schemes with $\Nt \in [16,128]$, $\Nr=\Nrf=\Ns = 2$, and SNR $=10$ dB.}
		\label{fig_comp_Nt}
	\end{figure}

\begin{figure}[t]
		\includegraphics[scale=0.58]{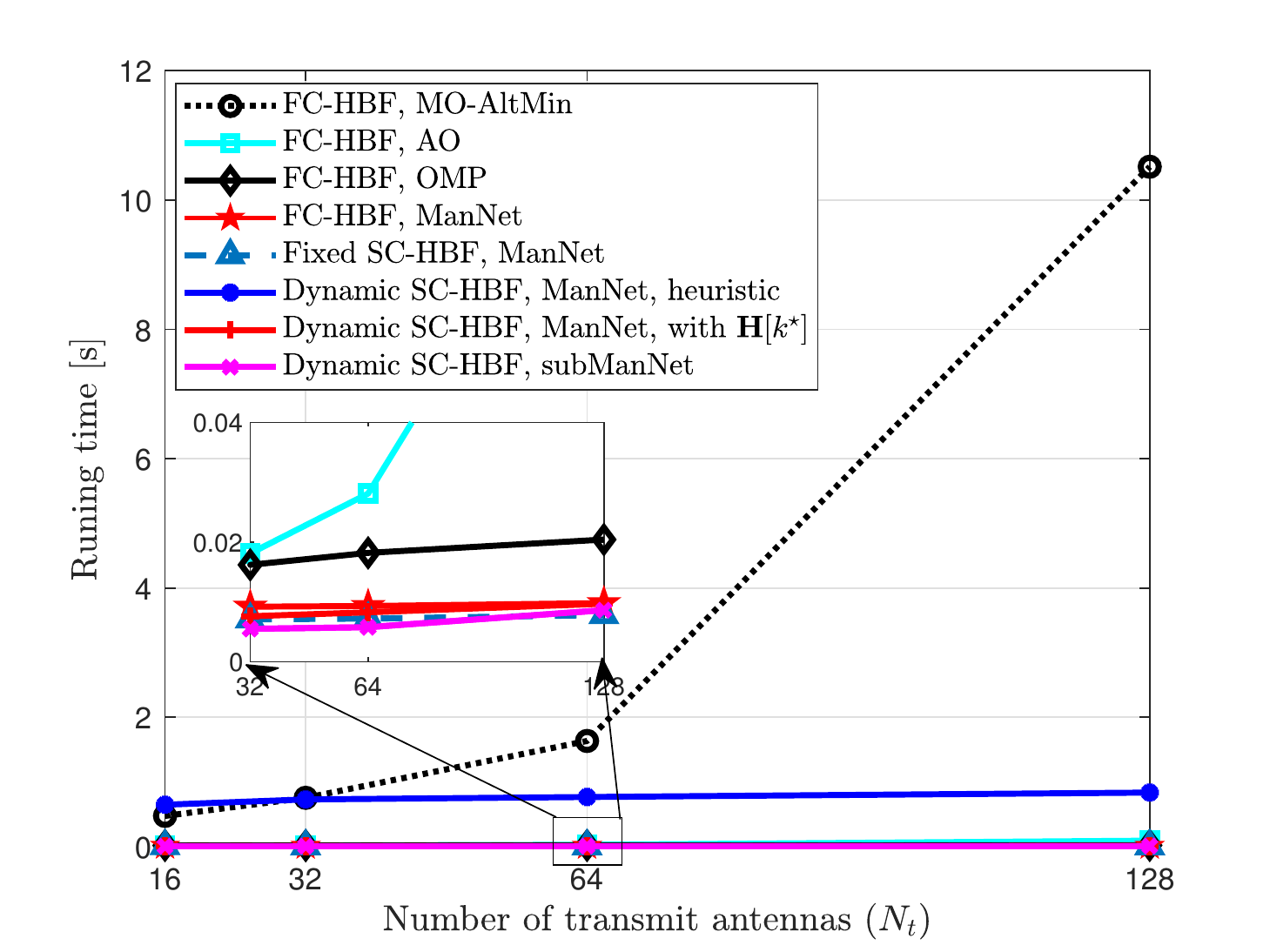}
		\caption{Run time of ManNet and subManNet-based HBF schemes with $\Nt \in [16,128]$, $\Nr=\Nrf=\Ns = 2$, and SNR $=10$ dB.}
		\label{fig_time_Nt}
	\end{figure}

	In Figs.\ \ref{fig_comp_Nt} and \ref{fig_time_Nt}, we compare the execution time and computational complexities of the considered schemes with the same simulation parameters as those for Fig.\ \ref{fig_SE_Nt}. The complexities are counted as the total number of additions and multiplications performed in the considered algorithms. The proposed deep unfolding schemes have low complexities thanks to ManNet and subManNet's small numbers of iterations, layers, and the simple operations in each layer. In particular, their complexities are just as low as OMP and slightly higher than SDR-AltMin, but they offer much better performance, as discussed earlier in Section \ref{sec_sim_performance}. Among the proposed deep unfolding schemes, as expected, subManNet-based SC-HBF has the lowest complexity, and the heuristic ManNet-based SC-HBF approach requires the highest complexity due to the iterations required for the search. Compared to these algorithms, the complexities of MO-AltMin and AO are much higher, and that of AO increases exponentially with $\Nt$, whereas the complexity of the algorithms is almost linear with $\Nt$. This agrees with the analysis in Section \ref{sec_complexity_sub}.
	
	Finally, we show the run time of the considered schemes in Fig.\ \ref{fig_time_Nt}, but we omit the results for SDR-AltMin because they are very large (up to $822$s for $\Nt=128$), making it difficult to see the difference among the other schemes. SDR-AltMin employs CVX to solve for the $\Fbb[k]$ in each iteration, and it is thus extremely slow. Among the other methods, MO-AltMin is the slowest and is much slower than AO, OMP, and the proposed deep unfolding approaches, especially for large $\Nt$. This is because of its slow convergence (see Fig.\ \ref{fig_rate_conv}) and nested iterations involving a line search. In contrast, the proposed deep unfolding algorithms execute very rapidly. With $\Nt = 128$, while MO-AltMin requires more than $10$s to execute, the time required by the non-heuristic ManNet and sub-ManNet-aided HBF schemes are only around $0.01$s. The heuristic ManNet-based dynamic SC-HBF approach outlined in Algorithm \ref{alg_ManNet_SC_HBF} requires a longer run time than the ManNet and subManNet-aided SC-HBF schemes. Furthermore, despite the slow convergence, AO executes relatively fast because only arithmetic operations and element-wise normalization are performed in each iteration.

	\section{Conclusion}
	\label{sec_conclusion}
	The nonconvexity and high-dimensional variables have imposed significant challenges to HBF designs in the literature. The available solutions have usually required cumbersome iterative procedures. We have overcome these difficulties by proposing efficient deep unfolding frameworks for FC-HBF and SC-HBF designs based on unfolding MO-AltMin and PGD. In these schemes, the low-complexity ManNet and subManNet approaches produce fully-connected and sub-connected analog precoders with only several layers and sparse connections in each, which explains the computational and time efficiency of the proposed algorithms. Our extensive simulation results demonstrate that compared to the state-of-the-art HBF algorithms, the proposed deep unfolding solutions for HBF designs have superior performance with lightweight implementation, low complexity, and fast execution. For future studies, deep unfolding models for a joint HBF design and channel estimation will be considered.

	\bibliographystyle{IEEEtran}
	\bibliography{IEEEabrv,Bibliography}

\end{document}